\documentclass[twocolumn,showpacs,preprintnumbers,amsmath,amssymb,unsortedaddress]{revtex4}

\usepackage{graphicx}  
\usepackage{dcolumn}   
\usepackage{bm}        

\def\lsim{~\rlap{$<$}{\lower 1.0ex\hbox{$\sim$}}}
\def\bsim{~\rlap{$>$}{\lower 1.0ex\hbox{$\sim$}}}
\def\kms{\ {\rm km\,s^{-1}}}

\def\hmpc{\ {\rm {\it h}^{-1}Mpc}}

\def\mdh{\ {\rm M_\odot/{\it h}}}

\def\keV{\,{\rm keV}}

\def\la{\langle}
\def\ra{\rangle}
\def\dd{{\rm d}}
\def\ln{{\rm ln}}
\def\tr{{\rm tr}}
\def\det{{\rm det}}

\def\mathbi#1{\textbf{\em #1}}

\def\dsc{\delta_{\rm sc}}

\def\rb{\bar{\rho}_m}
\def\ve{\mathbi{$\xi$}}

\def\rvh{\hat{\mathbi{r}}}

\def\rh{\hat{r}}

\def\vq{\mathbi{q}}
\def\vr{\mathbi{r}}
\def\vu{\mathbi{u}}
\def\vx{\mathbi{x}}
\def\vy{\mathbi{y}}
\def\ve{{\bf\eta}}
\def\vv{{\bf v}}

\def\vvs{{\bf S}}
\def\vaa{{\rm A}}
\def\vbb{{\rm B}}
\def\vcc{{\rm C}}
\def\vii{{\rm I}}
\def\vmm{{\rm M}}

\def\vrr{{\rm R}}

\def\vxx{{\rm X}}
\def\vyy{{\rm Y}}
\def\grad{\mathbi{$\nabla$}}

\def\hw{\hat{W}}
\def\dkk{\Delta_\delta^2(k)}

\def\mcur{\bar{u}}
\def\npk{n_{\rm pk}}
\def\vxpk{{\bf x}_{\rm pk}}
\def\vvpk{{\bf v}_{\rm pk}}
\def\gamu{\gamma_\upsilon}
\def\xpk{\xi_{\rm pk}}

\def\MNRAS{{Mon.~ Not.~ R.~ Astron.~ Soc.~}}

\def\PRD{{Phys.~ Rev.~ D.~}}
\def\PRL{{Phys.~ Rev.~ Lett.~}}
\def\ApJ{{Astrophys.~ J.~}}
\def\ApJS{{Astrophys.~ J.~ Suppl.~}}

\def\ApJSS{{Astrophys.~ J.~ Suppl.~ Ser.}}
\def\AA{{Astron.~ Astrophys.~}}
\def\Nat{{Nature (London)~}}
\def\AstroPart{{Astro-particle Phys.~}}

\def\JCAP{{JCAP}}
\def\BAMS{{Bull.~Am.~Math.~Soc.}}
\def\CMP{{Commun.~Math.~Phys.}}
\def\ApSS{{Astrophys.~Space~Science~}}
\def\MPLA{{Mod.~Phys.~Lett.~A}}
\def\PASJ{{Pub.~Astron.~Soc.~Jap.}}

\begin{document}

\title{Baryon acoustic signature in the clustering of density maxima}

\author{Vincent Desjacques
\small \\ \vspace{0.2cm} Institute for Theoretical Physics, 
University of Z\"urich, Winterthurerstrasse 190, CH-8057 Z\"urich, 
Switzerland \\
email: {\tt dvince@physik.uzh.ch}}


\begin{abstract}

We reexamine the two-point correlation of density maxima in Gaussian
initial conditions. Spatial derivatives of the linear density
correlation, which  were  ignored in the calculation of Bardeen et al.
[\ApJ, {\bf 304}, 15 (1986)],  are included in our analysis. These
functions exhibit large oscillations around the sound horizon scale
for generic CDM power  spectra. We derive the exact leading-order
expression for the correlation of density peaks and demonstrate the
contribution of those spatial derivatives.  In particular, we show
that these functions can modify significantly the baryon acoustic
signature of density maxima relative   to that of the linear density
field. The effect depends upon the  exact value of the peak height,
the filter shape and size,  and the small-scale behaviour of the
transfer function.  In the $\Lambda$CDM cosmology, for maxima
identified in the density field smoothed at mass scale  $M\approx
10^{12}-10^{14}\mdh$ and with linear threshold height
$\nu=1.673/\sigma(M)$, the contrast of the BAO can be a few tens of
percent larger than in the linear matter correlation. Overall, the BAO
is amplified for  $\nu\bsim 1$ and damped for $\nu\lsim 1$. Density
maxima thus behave  quite differently than linearly biased tracers of
the  density field, whose acoustic signature is a simple scaled
version of  the linear baryon acoustic oscillation. We also calculate
the mean streaming of peak pairs in the quasi-linear regime. We show
that the leading-order 2-point correlation and pairwise velocity of
density peaks are consistent with a nonlinear, local biasing relation
involving gradients of the density field.  Biasing will be an
important issue  in  ascertaining how much of the enhancement of the
BAO in the primeval correlation of density maxima propagates into the
late-time clustering of galaxies.

\end{abstract}

\maketitle

\section{Introduction}
\label{sec:intro}

Sound waves propagating in the primordial photon-baryon fluid imprint
a oscillatory  pattern in the anisotropies of the Cosmic Microwave
Background (CMB) and  in the matter distribution, whose characteristic
length scale $r_s$ is  the sound horizon at the recombination epoch
~\cite{baotheory}. $r_s\approx 105\hmpc$ for the  currently favoured
cosmological models. While experiments  have  accurately measured this
fundamental scale and its harmonic series in the temperature and
polarisation power spectra of the CMB, this acoustic signature has
recently been detected in the correlation function of galaxies
~\cite{Eisensteinetal2005,baoobs}. There is also weak evidence for the
baryon oscillations in the correlation function of clusters
~\cite{Estradaetal2008}.  In the 2-point correlation, the 
series of maxima and minima present in the  power spectrum translates
into a broad peak at the sound horizon scale. Since the latter can 
be accurately calibrated with CMB measurements, the baryon
acoustic oscillations (BAO) have emerged as a very promising standard
ruler for determining the   angular diameter distance and Hubble
parameter~\cite{baoprobe}.  Measuring the BAO at different redshifts
thus offers a potentially robust probe of the dark energy equation
of state.

In linear theory, the amplitude of the baryon acoustic peak increases
while its shape and  contrast remain unchanged.  However, the
clustering of galaxies  does not fully represent the primeval
correlation. Mode-coupling, pairwise velocities  and galaxy bias are
expected to alter the position and  shape of the acoustic peak and,
therefore, bias the measurement~\cite{3dbias}.  The evolution of the
acoustic pattern in the 2-point statistics of the matter, halo or
galaxy distributions has been studied using both numerical
simulations~\cite{Meiksinetal1999,baosimu} and  analytic techniques
based on the halo model or perturbation
theory~\cite{JeongKomatsu2006,SchultzWhite2006,
Guziketal2007,Eisensteinetal2007a,Smithetal2007,MatarresePietroni2007,
CrocceScoccimarro2008,Smithetal2008,Matsubara2008,MatarresePietroni2008,
Seoetal2008,JeongKomatsu2008}.  Yet the results of these studies do
not always agree  and the impact of nonlinearities on the matter and
galaxy power  spectrum remains debatable.  For instance,
References~\cite{SchultzWhite2006,Eisensteinetal2007a} argue that  any
systematic shift (i.e. not related to random motions or biasing) must
be less than the percent level owing to the particularly smooth  power
added by nonlinearities on those scale, and to the cancellation of the
mean streaming of (linearly) biased tracers at first order.  On the
other hand,
References~\cite{Guziketal2007,Smithetal2007,CrocceScoccimarro2008,
Smithetal2008} have shown that mode-coupling modifies the acoustic
pattern in the correlation of dark matter and haloes, and generates a
percent shift towards smaller scales. Despite their redshift
dependence, these shifts appear to be predictable and could be removed
from the data~\cite{Seoetal2008}.

There is a broad consensus regarding the shape of the acoustic
peak. In light of the nonlinear gravitational evolution of matter
fluctuations, it is sensible to expect a baryon acoustic peak  less
pronounced in the late-time  clustering  of galaxies than in the
linear theory correlation.   This can be shown to hold for any local
transformation of the density field
~\cite{Eisensteinetal2007a,Coles1993,ScherrerWeinberg1998}.  Such
biasing mechanisms do indeed predict a damping of the baryon acoustic
features in the 2-point statistics of the galaxy
distribution~\cite{Meiksinetal1999,SeoEisenstein2005,Smithetal2007}.
Galaxies, of course, form a discrete set of points but one commonly
assumes them to be a Poisson sample of some continuous field. Still,
the extent to which those models are an accurate approximation to the
clustering of galaxies remains unclear.   Notice also that a
reconstruction of the primordial density  field could significantly
restore the original contrast of the acoustic
oscillation~\cite{Eisensteinetal2007b}.

The main objective of this paper is to demonstrate that the BAO in
the correlation of tracers of the density field can be noticeably
modified if we consider local biasing relations more sophisticated
than local transformations of the density field
~\cite{Szalay1988,FryGaztanaga1993}.  To this purpose, we will examine
the clustering of density maxima in the initial cosmological density
field. In this respect, we will assume  that the initial fluctuations
are  described by Gaussian statistics. This assumption is remarkably
well supported by  measurements of the CMB and large-scale structures
~\cite{gaussianity,wmap5}.  Density peaks form a well-behaved
point-process whose statistical  properties depend not only on the
underlying density field, but also on its first and second
derivatives.  Therefore, although the number of density maxima is
modulated by large-scale fluctuations in the background, their
clustering properties cannot be derived from a continuous field
approach in which the peak overdensity would depend upon the value of
the matter density only. Interestingly however, we shall see that, at
large separations, the peak correlation and pairwise velocity are
consistent with a nonlinear biasing relation involving gradients of
the density field.

In a seminal paper, Bardeen, Bond, Kaiser \& Szalay (hereafter BBKS)
~\cite{Bardeenetal1986} provided  a compact expression for the average
number density of peaks in a three-dimensional Gaussian random field,
etc.  Furthermore, they obtained a large-scale approximation for the
correlation function of peaks which, at large threshold height, tends
toward the correlation of overdense
regions~\cite{Kaiser1984,PolitzerWise1984,JensenSzalay1986} as it
should be.  However, BBKS determined the correlation function of
density maxima only in the limit  where derivatives of the 2-point
function of the density field can be ignored. As we will see below,
these correlations  can greatly influence the large-scale correlation
of density maxima for generic Cold Dark Matter (CDM) power spectra.
It is also worth noticing that the statistics  of Gaussian random
fields in a cosmological context has received some attention in the
literature~\cite{Doroshkevich1970,HoffmanShaham1985,
PeacockHeavens1985,Coles1989,Lumsdenetal1989}. Some of these results
have been applied to the mass function and  correlation of rich
clusters for example~\cite{KaiserDavis1985,Ottoetal1986,Cen1998}.  The
present work mainly follows the analytic study of BBKS, and the lines
discussed in ~\cite{Desjacques2007,DesjacquesSmith2008}, where 2-point
statistics of the linear tidal shear are investigated. We refer the
reader to ~\cite{AdlerTaylor2007} for a rigorous introduction to the
statistics of maxima of Gaussian random fields.

The paper is organised as follows. Section ~\ref{sec:basics}
introduces a number of useful variables and correlation functions.
Section~\ref{sec:pkcorr} is devoted to the derivation of the leading
order expression for the large-scale asymptotics of the  peak
correlation. Our result can be thought as arising from a specific type
of nonlinear local biasing relation including second spatial
derivatives of the density field.   In Sec.~\ref{sec:pkclust}, we
explore the impact of these derivatives on the amplitude and shape of
the correlation of density maxima. Our attention  focuses on the
baryon oscillation, across  which the amplitude of the linear matter
correlation varies abruptly. It is shown that the  BAO of density
maxima can be amplified relative to that  of the matter
distribution. Section~\ref{sec:pkstream} deals with the peak pairwise
velocity. Its leading order contribution is found to be consistent
with the nonlinear local bias relation inferred from the 2-point
correlation of peaks.   A final section summarises our results.

\section{Properties of cosmological Gaussian density fields}
\label{sec:basics}

We review some general properties of Gaussian random fields and
provide  explicit expressions for the correlations of the density and
its lowest  derivatives. We show that the latter are not always 
negligible in CDM cosmologies.

\subsection{Useful definitions}

We will assume a $\Lambda$CDM cosmology with normalisation amplitude
$\sigma_8=0.82$, and spectral index $n_s=0.96$~\cite{wmap5}.  The
matter transfer function is computed using publicly available
Boltzmann codes~\cite{boltzmann}. The position of  the BAO in the
linear matter correlation function is close  to $\approx
105.0\hmpc$. 

Let $\vq$ designate the Lagrangian coordinate.  We are interested in
the three-dimensional density field $\delta(\vq)$ and its first and
second derivatives.  It is more convenient to work with the normalised
variables $\nu=\delta(\vq)/\sigma_0$,
$\eta_i=\partial_i\delta(\vq)/\sigma_1$ and
$\zeta_{ij}=\partial_i\partial_j\delta(\vq)/\sigma_2$, where the
$\sigma_j$  are the spectral moments of the matter power spectrum,
\begin{equation}
\sigma_j^2 \equiv \int_0^\infty\!\!\dd\ln k\,k^{2j}\,\Delta^2(k)\;.
\label{eq:mspec}
\end{equation} 
$\Delta^2(k)\equiv \dkk|\hw(k,R_f)|^2$ denotes the dimensionless
power spectrum of the density field smoothed on scale $R_f$ with a
spherically symmetric window $\hw(k,R_f)$. 

The best choice of smoothing is open to debate. Among the popular
window functions, the top hat filter is compactly supported and has a
straightforward interpretation within the spherical collapse
model. Notwithstanding this, oscillations that arise in Fourier space
do not lead to well defined spectral moments $\sigma_j$ with $j\geq 2$
for CDM power spectra. This can be understood by examining the
high-$k$ tail of the CDM transfer function.  Neglecting the baryon
thermal pressure on  scale less than the Jeans length, the small-scale
matter transfer  function behaves as $T(k)\propto
\ln(1.8k)/k^2$~\cite{Bardeenetal1986,tksmalltail}, which clearly leads
to divergences when the integer $j$ is larger than one. By contrast, a
Gaussian window function ensures the convergence of all the spectral
moments for any realistic matter power spectra. Consequently, we shall
mostly rely on the Gaussian filter throughout this paper, although the
top hat filter will also be considered briefly in Sec.
~\ref{sec:pkclust}. Note that a Gaussian filter of characteristic
width $R_f$ encloses  a  mass $M_f=(2\pi)^{3/2}\bar{\rho} R_f^3$ a few
times larger than that encompassed by a top hat filter of identical
smoothing radius.

Following BBKS, we also introduce the parameters
$\gamma=\sigma_1^2/(\sigma_0\sigma_2)$  and
$R_\star=\sqrt{3}\sigma_1/\sigma_2$ for subsequent use. The spectral
width $\gamma$ reflects  the range over which $\Delta^2(k)$ is large,
while $R_\star$  characterises the radius of peaks. For the special
case of a powerlaw power spectrum with Gaussian filtering on scale
$R_f$, these parameters are given by $\gamma^2=(n+3)/(n+5)$ and
$R_\star^2=6R_f^2/(n+5)$. For CDM power spectra, $\gamma\sim 0.5-0.7$
when the smoothing length varies over the range  $0.1\lsim R_f\lsim
10\hmpc$.

\begin{figure}
\resizebox{0.45\textwidth}{!}{\includegraphics{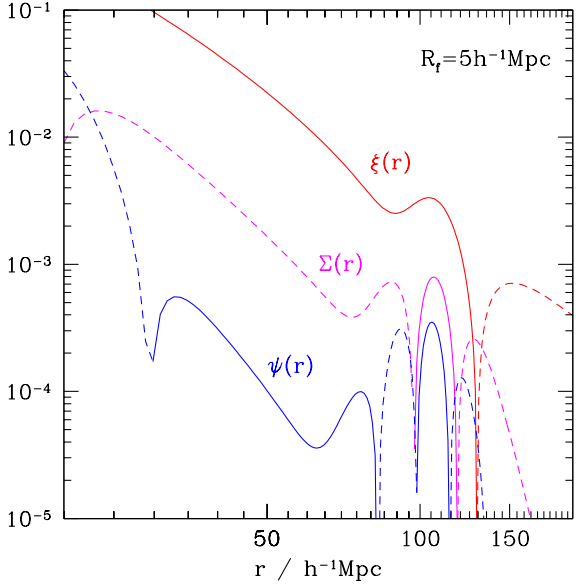}}
\caption{A comparison between the cross-correlation of the density
field, $\xi(r)$, and that of its first and second derivatives,
$\Sigma(r)$ and $\psi(r)$ respectively (see
eq.~\ref{eq:spaderivs}). Results are shown as a function of the
Lagrangian  separation $r$ for the $\Lambda$CDM cosmology considered
in the present work. The density field is smoothed with a Gaussian
filter of characteristic scale $R_f=5\hmpc$ (i.e. a mass scale
$M_f=1.5\times 10^{14}\mdh$).  Dashed lines denote negative values.
All the correlations are normalised to unity at zero lag.}
\label{fig:xir1}
\end{figure}

\subsection{Correlation of the density and its derivatives}

Calculating the 2-point correlation of density peaks requires knowledge
of the auto- and cross-correlations of the various fields. These objects
can be decomposed into components with definite transformation properties 
under rotations. Statistical isotropy and symmetry implies that, in 
position space, the most general ansatz for the isotropic sector of 
the 2-point correlations of these fields reads
\begin{eqnarray}
\lefteqn{\la\nu(\vq_1)\nu(\vq_2)\ra = \xi(r)} && \\
\lefteqn{\la\nu(\vq_1)\eta_i(\vq_2)\ra = \Xi(r)\,\rh_i} && 
\nonumber \\
\lefteqn{\la\nu(\vq_1)\zeta_{ij}(\vq_2)\ra = -\gamma\Sigma_1(r)
\,\rh_i\rh_j - \gamma\Sigma_2(r)\,\delta_{ij}} && \nonumber \\
\lefteqn{\la\eta_i(\vq_1)\eta_j(\vq_2)\ra = \Sigma_1(r)\,\rh_i\rh_j +
\Sigma_2(r)\,\delta_{ij}} && \nonumber \\
\lefteqn{\la\eta_i(\vq_1)\zeta_{lm}(\vq_2)\ra} && \nonumber \\  
&& =\Pi_1(r)\,\rh_i\rh_l\rh_m +\Pi_2(r)\,\left(\rh_i\delta_{lm} 
+ \rh_l\delta_{im} + \rh_m\delta_{il}\right) \nonumber \\
\lefteqn{\la \zeta_{ij}(\vq_1)\zeta_{lm}(\vq_2)\ra = \Psi_1(r)\,\rh_i
\rh_j \rh_l \rh_m} \nonumber \\ && +\,\Psi_3(r)\left(\rh_i
\rh_l\delta_{jm}+\rh_i \rh_m\delta_{jl} +\rh_j \rh_l\delta_{im}+\rh_j
\rh_m\delta_{il}\right. \nonumber \\ && \left.\rh_i\rh_j\delta_{lm}
+\rh_l \rh_m\delta_{ij}\right)+\Psi_5(r)\,\left(\delta_{ij}\delta_{lm}
+\delta_{il}\delta_{jm}+\delta_{im}\delta_{jl}\right) \nonumber \;,
\label{eq:correl}
\end{eqnarray}
where $r=|\vq_2-\vq_1|$ is the Lagrangian separation, $\rh_i=r_i/r$
and the functions $\xi$, $\Xi$,  $\Sigma_i$, $\Pi_i$ and $\Psi_i$
depend on $r$ only. We emphasise that these correlation functions
transform as scalar under rotations. Note also that these expressions
are valid for any arbitrary random field.  For a cosmological Gaussian
density field however, these functions can be  summarised as follows~:
\begin{eqnarray}
\xi(r)\!\!\! &=& \!\!\! \frac{1}{\sigma_0^2}\int_0^\infty\!\!\dd\ln k\,
\Delta^2(k)\, j_0(kr) \\
\Xi(r)\!\!\! &=& \!\!\! -\frac{1}{\sigma_0\sigma_1}\int_0^\infty\!\!
\dd\ln k\,k\Delta^2(k)\, j_1(kr) \nonumber \\
\Sigma_1(r)\!\!\! &=& \!\!\! -\frac{1}{\sigma_1^2}\int_0^\infty\!\!
\dd\ln k\, k^2\Delta^2(k)\,j_2(kr) \nonumber \\
\Sigma_2(r)\!\!\! &=& \!\!\! \frac{1}{\sigma_1^2}\int_0^\infty\!\!
\dd\ln k\, k^2\Delta^2(k)\,\left[\frac{1}{3}j_0(kr)+\frac{1}{3}j_2(kr)
\right] \nonumber \\
\Pi_1(r)\!\!\! &=& \!\!\! -\frac{1}{\sigma_1\sigma_2}\int_0^\infty\!\!
\dd\ln k\, k^3\Delta^2(k)\,j_3(kr) \nonumber \\
\Pi_2(r)\!\!\! &=& \!\!\! \frac{1}{\sigma_1\sigma_2}\int_0^\infty\!\!
\dd\ln k\, k^3\Delta^2(k)\,\left[\frac{1}{5}j_1(kr)+\frac{1}{5}j_3(kr)
\right] \nonumber \\
\Psi_1(r)\!\!\! &=& \!\!\! \frac{1}{\sigma_2^2}\int_0^\infty\!\!\dd\ln
k\,k^4\Delta^2(k)\, j_4(kr) \nonumber \\ 
\Psi_3(r) \!\!\! &=& \!\!\! -\frac{1}{\sigma_2^2}\int_0^\infty\!\!
\dd\ln k\,k^4\Delta^2(k)\left[\frac{1}{7}j_2(kr)+
\frac{1}{7}j_4(kr)\right] \nonumber \\
\Psi_5(r) \!\!\! &=& \!\!\!\frac{1}{\sigma_2^2}\int_0^\infty\!\!\dd\ln k
\,k^4\Delta^2(k) \nonumber \\ && \times
\left[\frac{1}{15}j_0(kr)+\frac{2}{21}j_2(kr)+\frac{1}{35}j_4(kr)\right] 
\;. \nonumber
\label{eq:psif}
\end{eqnarray}
$j_\ell(x)$ are spherical Bessel functions of the first kind. In the 
limit $r\rightarrow 0$, all the correlation functions vanish but $\xi$, 
$\Sigma_2$ and $\Psi_5$, which tend towards 1, 1/3 and 1/15, respectively.
Averaging over the direction $\rvh$ of the separation vector thus yields
\begin{eqnarray}
\frac{1}{4\pi}\!\!\int\!\!\dd\Omega_{\rvh}\,\la\eta_i(\vq_1)
\eta_j(\vq_2)\ra \!\!\!&=&\!\!\! \frac{\Sigma(r)}{3}\,\delta_{ij} \\
\frac{1}{4\pi}\!\!\int\!\!\dd\Omega_{\rvh}\,\la\zeta_{ij}(\vq_1)
\zeta_{lm}(\vq_2)\ra \!\!\! &=& \!\!\! \frac{\psi(r)}{15}\,
\left(\delta_{ij}\delta_{lm}+\delta_{il}\delta_{jm}
+\delta_{im}\delta_{jl}\right)\nonumber
\end{eqnarray}
for the covariances of the fields $\eta_i$ and $\zeta_{ij}$, where we 
have defined
\begin{eqnarray}
\Sigma(r) &=& \Sigma_1(r)+3\Sigma_2(r) \\
\psi(r) &=& \Psi_1(r)+10\Psi_3(r)+15\Psi_5(r) \nonumber \;.
\label{eq:spaderivs}
\end{eqnarray}
The angular average of the other correlation functions vanishes,
except that of the density correlation of course. 

$\Sigma(r)$ and $\psi(r)$ can be  expressed in terms of the
derivatives of the density correlation  using relations like
$\la\eta_i\eta_j\ra=-\partial_i\partial_j\xi(r)$  etc. For a density
correlation that falls off as a powerlaw $r^{-n-3}$,  $\Sigma(r)$ and
$\psi(r)$ decay as $r^{-n-5}$ and $r^{-n-7}$,  respectively. This
derivation assumes a powerlaw power spectrum with a fair amount of
power at short wavenumbers. Hence, as recognised
in BBKS, neglecting the derivatives of the density
correlation should be a reasonable approximation when $n\lsim -1$.

\begin{figure}
\resizebox{0.45\textwidth}{!}{\includegraphics{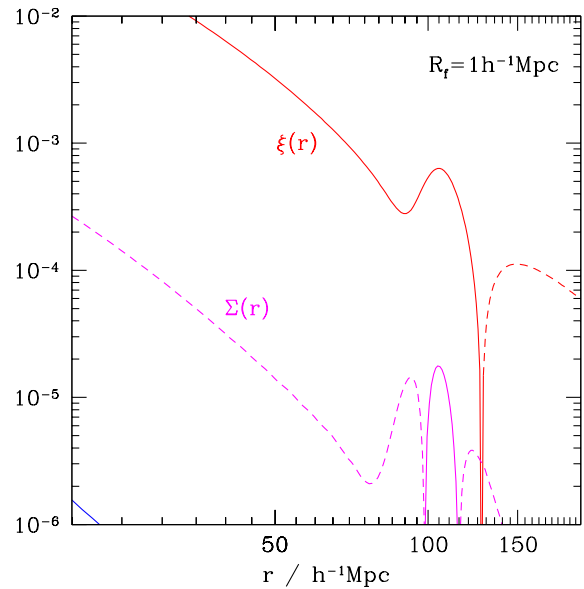}}
\caption{Same as Fig.~\ref{fig:xir1}, but for a smoothing length
$R_f=1\hmpc$ ($M_f=1.2\times 10^{12}\mdh$). The correlation function
$\psi(r)$ (not shown) is less than $10^{-6}$ at distances larger than
$\bsim 30\hmpc$.}
\label{fig:xir2}
\end{figure}

This simple argument may not hold for CDM cosmologies since the index
$n$ is a smooth function of the separation $r$.  Namely, it is $n\sim
-2$ when $r\sim 10\hmpc$, and increases to attain a value of the order
of unity on scale $r\bsim 60\hmpc$.  For illustration purpose, the
functions $\xi$, $\Sigma$ and $\psi$ are plotted in Figures 
\ref{fig:xir1} and \ref{fig:xir2}
for the $\Lambda$CDM cosmology considered here. The filtering length
is $R_f=5$ and $1\hmpc$, respectively (The reason for choosing these
values will become apparent below).  Retaining only
the density correlation appears to be a good  approximation on scales
larger than a few smoothing radii. However, the relative amplitude of
the cross-correlations strongly depends upon the filtering scale.
Namely, both $\Sigma(r)$ and $\psi(r)$ increase relative to $\xi(r)$ 
with increasing smoothing length. Yet another striking feature of
Figures~\ref{fig:xir1} and \ref{fig:xir2} 
is the oscillatory behaviour of $\Sigma(r)$ and
$\psi(r)$. The large  oscillations are caused by rapid changes in
the linear matter correlation across the baryon acoustic peak. Notice
that both $\Sigma(r)$ and $\psi(r)$ are positive at distances
$r\approx 100-110\hmpc$. On  these scales, when $R_f=1\hmpc$,
$\Sigma(r)$ reaches  to 3 per cent of the density correlation while
$\psi(r)$ is negligible.  At the large smoothing length however,
they nearly  reach 20 and 10 per cent of the density correlation, 
respectively.

These results suggest that, for generic CDM power spectra, the
derivatives of the density  correlation could have a significant
impact on the correlation of density maxima, especially in the
vicinity of the baryon acoustic feature. This motivates the 
calculation presented in the next Section.

\section{Correlation of density maxima}
\label{sec:pkcorr}

Owing to the constraints on the derivatives of the density field,
calculating the $n$-point correlation function of  peaks requires
performing integration over a joint probability  distribution in 10$n$
variables. Therefore, even the evaluation of the 2-point correlation
of density maxima $\xpk(r)$ proves difficult.   Here, we derive the
leading order expression that includes, in addition  to the linear
matter correlation $\xi(r)$, the  contribution of the angular average
functions $\Sigma(r)$ and $\psi(r)$. We also show that the large-scale 
asymptotics of the peak correlation can be thought as arising from 
a specific type of nonlinear biasing relation involving second 
derivatives of the density field.

\subsection{The Kac-Rice formula}

As shown in BBKS for instance, the correlation of density extrema
(maxima, minima and saddle points) can be entirely expressed in terms
of $\delta(\vq)$ and its derivatives, $\eta_i(\vq)$ and
$\zeta_{ij}(\vq)$.  In the neighbourhood of an extremum, the first
derivative $\eta_i$ is approximately
\begin{equation}
\eta_i(\vq)\approx\sqrt{3}R_\star^{-1}\,\sum_j\zeta_{ij}(\vq_p)
\left(\vq-\vq_p\right)\;.
\end{equation}
Using the properties of the Dirac delta, the number density of extrema 
can be written as 
\begin{equation}
n_{\rm ext}(\vq)=\sum_p \delta^3\!\left(\vq-\vq_p\right)=
\frac{3^{3/2}}{R_\star^3}\,|\det\zeta(\vq)|
\delta^3\!\left[\ve(\vq)\right]\;,
\label{eq:next}
\end{equation}
provided that the Hessian $\zeta_{ij}$ is invertible. The delta
function $\delta^3[\ve]$  ensures that all the extrema are
included. In this paper however, we are  interested in counting the
maxima solely. Consequently, we further have  to require
$\zeta_{ij}(\vq_p)$ be negative definite at the extremum  position
$\vq_p$. Note that, later, we will also restrict the set to  those
maxima with a certain threshold height. The average number density  of
maxima eventually reads
\begin{equation}
\la\npk(\vq)\ra=\frac{3^{3/2}}{R_\star^3}\,
\la|\det\zeta(\vq)|\delta^3\!\left[\ve(\vq)\right]\ra\;. 
\label{eq:kacrice}
\end{equation}
This expression, known as the Kac-Rice formula 
~\cite{Kac1943,Rice1954,Belyaev1967,Clineetal1987,AdlerTaylor2007}, 
holds for arbitrary smooth random fields. In the general case of a 
random field in $N$ dimensions, it is trivial to show that the mean 
density of maxima scales as 
$\la\npk\ra\propto R_\star^{-N}\propto R_f^{-N}$.

The 2-point correlation function of density peak is defined such that
\begin{equation}
1+\xpk(\vr)=\la\npk(\vq_1)\npk(\vq_2)\ra/\la\npk\ra^2
\label{eq:2pk1}
\end{equation}
is the joint probability that a density maxima is in a volume  $\dd
V_i$ about each $\vq_i$.  Let $\vxx$ be the diagonal matrix of entries
${\rm diag}(x_1,x_2,x_3)$ where  $x_1\geq x_2\geq x_3$ is the
non-increasing sequence of eigenvalues of the symmetric matrix
$-\zeta$. The condition that the extrema are maxima  implies $x_3\geq
0$. Therefore, the correlation function of peaks is given by
\begin{eqnarray}
\lefteqn{1+\xpk(\vr)} &&    \\ &&
=\frac{3^3}{\la\npk\ra^2 R_\star^6}\,
\la|\det\zeta_1||\det\zeta_2|\,\theta(x_3)
\theta(y_3)\,\delta^3\![\ve_1]\delta^3\![\ve_2]\ra \nonumber \\ &&
=\frac{3^3}{\la\npk\ra^2 R_\star^6}
\int\!\!\dd\nu_1\dd^6\zeta_1\dd\nu_2
\dd^6\zeta_2\,|\det\zeta_1||\det\zeta_2|\,\theta(x_3)\theta(y_3)
\nonumber \\  && \hspace{5mm} \times\,
P\left(\ve_1=0,\nu_1,\zeta_1,\ve_2=0,\nu_2,\zeta_2;\vr\right)\nonumber\;,
\label{eq:2pk2}
\end{eqnarray}
where, for shorthand convenience, subscripts denote quantities
evaluated at different Lagrangian positions, and $\dd^6\zeta=\prod_{i\leq
j}\dd\zeta_{ij}$ is the usual Lebesgue measure on  the six-dimensional
space of symmetric matrices. Here and  henceforth $\theta(x)$
designates the Heaviside step-function, i.e.  $\theta(x)=1$ for $x>0$
and zero otherwise.

\subsection{Two-point probability distribution}

The joint probability  distribution of the density fields together
with its first and second derivatives, 
$P\left(\ve_1,\nu_1,\zeta_1,\ve_2,\nu_2,\zeta_2;\vr\right)$, is
given by a multivariate Gaussian whose covariance matrix $\vcc$ has 20
dimensions.   This $20 \times 20$ matrix may be partitioned into four
$10\times 10$ block matrices,
$\vmm=\la\vy_1\vy_1^\top\ra=\la\vy_2\vy_2^\top\ra$ in  the top  left
corner and bottom right corners,  $\vbb=\la\vy_1\vy_2^\top\ra$ and its
transpose in the  bottom left and  top right corners,
respectively. The components $\zeta_A$, $A=1,\dots,6$  of the
ten-dimensional vector $\vy^\top=(\eta_i,\nu,\zeta_A)$ symbolise the
entries  $ij=11,22,33,12,13,23$ of $\zeta_{ij}$. To emphasise that 
the entries $\zeta_A$ transform as a tensor under rotation, we shall 
also label them as the matrix $\zeta$ in what follows. 

The matrices $\vmm$ and  $\vbb$ can be further decomposed into block 
sub-matrices of size 4 and 6,
\begin{equation}
\vmm=\left(\begin{array}{cc}\vmm_1 & \vmm_3^\top \\  \vmm_3 &
\vmm_2\end{array}\right),~~~ \vbb=\left(\begin{array}{cc} \vbb_1 &
\vbb_3^\top \\ \vbb_3 & \vbb_2 \end{array}\right) \;.
\end{equation} 
Unlike the $\vmm_i$ which describe the covariances at a single position, 
the matrices $\vbb_i$ generally are functions of the separation vector
$\vr$. Using the harmonic decomposition of the tensor products 
$\rvh\otimes\dots\otimes\rvh$, they can be written as
\begin{eqnarray}
\vbb_1(\vr)&=& \vbb_1^{0,0}+\sum_{\ell=1}^4
\vbb_1^{\ell,m}(r)\,Y_\ell^m(\rvh) \nonumber \\
\vbb_2(\vr)&=& \vbb_2^{0,0}+\sum_{\ell=1}^4
\vbb_2^{\ell,m}(r)\,Y_\ell^m(\rvh) \nonumber \\
\vbb_3(\vr)&=& \vbb_3^{0,0}+\sum_{\ell=1}^4
\vbb_3^{\ell,m}(r)\,Y_\ell^m(\rvh)\;.
\label{eq:multipole}
\end{eqnarray}
$Y_\ell^m(\rvh)$ are spherical harmonics and $\vbb_i^{\ell,m}(r)$ are 
matrices which satisfy $(\vbb_i^{\ell,m})^\dagger=(-1)^m\vbb_i^{\ell,m}$. 
Only multipoles up to $\ell=4$ appear in the harmonic decomposition since
the correlations given in eq.~(\ref{eq:correl}) involve products of up to 
four unit vectors $\rh$. The monopole terms are
\begin{eqnarray}
\vbb_1^{0,0} &=& \left(\begin{array}{cc}\Sigma(r)/3\,\vii & 0_{3\times 1} \\  
0_{1\times 3} & \xi(r)\end{array}\right) \;, \\ 
\vbb_2^{0,0} &=& \left(\begin{array}{cc} \psi(r)/15\,\vaa & 0_{3\times 3} \\
0_{3\times 3} & \psi(r)/15\,\vii\end{array}\right) \;, \nonumber \\
\vbb_3^{0,0} &=& \left(\begin{array}{cc} 0_{3\times 3} & 
-\gamma\Sigma(r)/3\,1_{3\times 1} \\ 
0_{3\times 3} & 0_{3\times 1}\end{array}\right)\;,
\nonumber
\label{eq:monopole}
\end{eqnarray}
where 
\begin{equation}
\vaa=\left(\begin{array}{ccc} 3&1&1 \\ 1&3&1 \\ 1&1&3\end{array}\right)\;,
\end{equation} 
$\vii$ is the $3\times 3$ identity matrix and $1_{1\times 3}=(1,1,1)$ etc.
The matrices $\vmm_i$ are readily obtained as $\vmm_i=\vbb_i^{0,0}(0)$.
An explicit computation of the higher multipole matrices is unnecessary 
here as we confine the calculation to the monopole contribution.

It is important to note that the joint density $P(\vy_1,\vy_2,\vr)$
preserves its functional form under the action of the rotation group
SO(3).  However, in a given frame of  reference, $P(\vy_1,\vy_2,\vr)$
does change when $\rh$ moves on the unit sphere. Does this mean that
$\xpk(\vr)$ truly depends on the direction of the separation vector
$\vr$ ? No, as it should be clear from  eq.~(\ref{eq:2pk2}) where the
volume measure $|\det\zeta|\dd^6\zeta$ is a rotational invariant.
More precisely, the volume element $\dd^6\zeta$ can be cast into the 
form
\begin{equation}
\dd^6\zeta=8\pi^2\,|\Delta (x)|\,\dd^3 x\,\dd\vrr\;.
\label{eq:vol1}
\end{equation}
where the $x_i$s are, as before, the three ordered eigenvalues of $-\zeta$, 
$\dd^3 x=\dd x_1\dd x_2\dd x_3$ and $\Delta(x)=\prod_{i<j}(x_i-x_j)$ is 
the Vandermonde determinant. $\dd\vrr$ is the Haar measure (for the
Euler angles for example) on the group SO(3) normalised to 
$\int\dd\vrr=1$. The peak correlation thus is proportional to
\begin{equation}
\int\dd\vrr_1\dd\vrr_2\,
P(\ve_1=0,\nu_1,\zeta_1,\ve_2=0,\nu_2,\zeta_2,\vr)\;,
\label{eq:average1}
\end{equation}
where the integral runs over the two SO(3) manifolds that define the
orientation of the principal frames of $\zeta_1$ and $\zeta_2$
relative to the frame of reference. Alternatively, we can choose the
coordinate system  such that the coordinate axes are aligned with the
principal axes of $\zeta_1$. In this new coordinate frame, the above
integral becomes
\begin{equation}
\frac{1}{4\pi}\int\!\!\dd\Omega_{\rvh}\,\dd\vrr\,
P(\ve_1=0,\nu_1,\zeta_1,\ve_2=0,\nu_2,\zeta_2,\vr)\;,
\label{eq:average2}
\end{equation}
where $\vrr$ is an orthogonal matrix that defines the orientation
of the eigenvectors of $\zeta_2$ relative to those of $\zeta_1$. This
demonstrates that only the monopole component of $P(\vy_1,\vy_2,\vr)$
contributes to the peak correlation function. Therefore, $\xpk(\vr)$
is invariant under rotations of the coordinate system, namely, it is 
a function of the separation $r$ only.

\subsection{Large scale asymptotics}

To obtain the correlation function of peak, we need first to calculate 
the 2-point probability distribution function averaged over the unit 
sphere for the variables $\vy^\top=(\eta_i,\nu,\zeta_A)$,
\begin{equation}
P(\vy_1,\vy_2,r)=\frac{1}{4\pi}\int\!\!\dd\Omega_{\rvh}\,
P(\vy_1,\vy_2,\vr)\;.
\end{equation}
In the large-distance limit ($r\gg 1$), the cross-correlation matrix 
is small when compared to the zero-point contribution $\vmm$, e.g.
$|\vbb|\ll\vmm$.  Following~\cite{Desjacques2007,DesjacquesSmith2008},
the quadratic form  which appears in the probability distribution 
$P(\vy_1,\vy_2;\vr)$,
\begin{equation}
P(\vy_1,\vy_2;\vr)=\frac{1}{\left(2\pi\right)^{10}|\det\vcc|^{1/2}}\,
e^{-Q(\vy_1,\vy_2,\vr)}\;,
\end{equation}
where $\det\vcc\approx|\det\vmm|^2=4^2\,(1-\gamma^2)^2/(15^{10}\,3^8)$
is the determinant of the covariance matrix $\vcc$, can be computed
easily using Schur's identities. Expanding the exponential  in the
small perturbation $\vbb$ yields, to first order,
\begin{equation}
e^{-Q(\vy_1,\vy_2,\vr)}\approx \left(1+\vy_1^\top\vmm^{-1}
\vbb\,\vmm^{-1}\vy_2\right)\,e^{-\bar{Q}(\vy_1,\vy_2)}\;,
\end{equation}
where the quadratic form $\bar{Q}(\vy_1,\vy_2)$ can be recast as
\begin{equation}
2\bar{Q}=\nu_1^2+\frac{\left(\gamma\nu_1+\tr\zeta_1\right)^2}{1-\gamma^2}
+\frac{5}{2}\left[3\tr(\zeta_1^2)-\left(\tr\zeta_1\right)^2\right]
+ 1\leftrightarrow 2\;,
\label{eq:qform1}
\end{equation}
in agreement with the results of BBKS. The calculation 
of $\vy_1^\top\vmm^{-1}\vbb\vmm^{-1}\vy_2$ is tedious but straightforward. 
Fortunately, only the monopole terms $\vbb_i^{0,0}$ survive after averaging 
over the directions $\rvh$. After further simplification, the result can
be reduced to the following compact expression~:
\begin{widetext}
\begin{eqnarray}
\lefteqn{\frac{1}{4\pi}\int\!\!\dd\Omega_{\rvh}\,\vy_1^\top\vmm^{-1}
\vbb\,\vmm^{-1}\vy_2 = \frac{5}{2}\left[3\,\tr\left(\zeta_1\zeta_2\right)
-\tr\zeta_1\,\tr\zeta_2\right]\psi(r)+\left\{\tr\zeta_1\,\tr\zeta_2
\left[\psi(r)+\gamma^2\xi(r)\right] \right.} && \\ && 
+ \left. \nu_1\nu_2\left[\xi(r)+\gamma^2\psi(r)\right]
-2\gamma^2\left(\tr\zeta_1\,\tr\zeta_2+\nu_1\nu_2\right)\Sigma(r)
+\gamma\left(\nu_1\tr\zeta_2+\nu_2\tr\zeta_1\right)
\left[\xi(r)+\psi(r)-\left(1+\gamma^2\right)\Sigma(r)\right]\right\} 
\left(1-\gamma^2\right)^{-2}\nonumber\;.
\label{eq:average3}
\end{eqnarray}
\end{widetext}
The invariance under rotation requires that
$P(\vy_1,\vy_2,r)$ be a  symmetric function of the eigenvalues, and
thus a function of  $\tr\left(\zeta_1^k\zeta_2^l\right)$, 
$k,l=0,1,\dots$

Since the above expression depends only upon the relative orientation
of the  two principal axes frames of $\zeta_1$ and $\zeta_2$ (through
the presence of $\tr(\zeta_1\zeta_2)$), we choose a coordinate system
whose  axes are aligned with the principal frame of
$\zeta_1$. With this choice of coordinate, we define $\zeta_1=-\vxx$
and $\zeta_2=-\vrr\vyy\vrr^\top$, where $\vrr$ is an orthogonal matrix
that defines the relative orientation of the eigenvectors of $\zeta_2$. 
$\vxx$ and $\vyy$ are the diagonal matrices
consisting of the three ordered eigenvalues $x_i$ and $y_i$ of the
Hessian $-\partial_i\partial_j\nu$. The properties of  the trace imply
that $\tr\zeta_1=-\tr\vxx$, $\tr(\zeta_1^2)=\tr(\vxx^2)$ (and
similarly for $\zeta_2$), while the term
$\tr(\zeta_1\zeta_2)=\tr(\vxx\vrr\vyy\vrr^\top)$ depends explicitely
on the rotation matrix $\vrr$.

The integral over the SO(3) manifold that describes the orientation of
the orthonormal triad of $\zeta_1$ is immediate. The result is $2\pi^2$
(and not $8\pi^2$) as we don't care whether the axes are directed
towards  the positive or negative direction. The integral over the
second SO(3)  manifold involves
\begin{equation}
\int_{{\rm SO(3)}}\!\!\!\!\!\!\dd\vrr\,\tr\left(\vxx\vrr\vyy\vrr^\top
\right)=\frac{1}{3}\,\tr\vxx\,\tr\vyy\;,
\end{equation}
and yields cancellation of the first term in the right-hand side of
eq.~(\ref{eq:average2}). To integrate over the eigenvalues of $\zeta_1$
and $\zeta_2$, we transform to the new set of variables
$\left\{u_i,v_i,w_i,i=1,2\right\}$, where
\begin{eqnarray}
&& u_1 = x_1+x_2+x_3 \nonumber \\ && v_1 = \left(x_1-x_3\right)/2
\nonumber \\  && w_1 = \left(x_1-2 x_2+x_3\right)/2 \;.
\label{eq:newset}
\end{eqnarray}
The variables $(u_2,v_2,w_2)$ are similarly defined in terms of the
$y_i$. We will henceforth refer to $u$ as the peak curvature.

Our choice of ordering imposes the constraints $v_i\geq 0$ and
$-v_i\leq w_i\leq v_i$. The condition that the density extrema be
maxima, i.e. all three eigenvalues of the Hessian 
$\zeta_{ij}$ are negative, translates into $(u_i+w_i)\geq 3v_i$.
Another condition, $u_i\geq 0$, should also be applied if one is
interested in selecting maxima with positive threshold  height.

For shorthand convenience, and to facilitate the comparison with the 
calculation of BBKS, we introduce the auxiliary 
function
\begin{eqnarray}
\lefteqn{F(u_1,v_1,w_1) \equiv \frac{3^3}{2}\,|\det\vxx|\,\Delta(x)} 
\\ && = \left(u_1-2w_1\right)\left[\left(u_1+w_1\right)^2-9v_1^2\right]
v_1\left(v_1^2-w_1^2\right)\nonumber \;,
\end{eqnarray}
$F(u_1,v_1,w_1)$  measures the degree of asphericity expected for a
peak and can be used to  determine the probability distribution of
ellipticity $v_1/u_1$ and prolateness $w_1/u_1$~\cite{Bardeenetal1986}.  
It scales as $\propto u_i^3$ in the limit $u_i\gg 1$.

\subsection{The peak correlation $\xpk(\nu,r)$}

For sake of generality, we will present results for the
cross-\-correlation  $\xpk(\nu_1,\nu_2,r)$ between two populations of
density maxima  $\nu_1\ne \nu_2$ identified at smoothing scale $R_1\ne
R_2$. However,  we shall focus shortly on the auto-correlation
$\xpk(\nu,r)$ ($\nu_1=\nu_2=\nu$ and $R_1=R_2=R_f$), which is more
directly related to the clustering properties of dark matter haloes of
a given mass or galaxies and clusters of a given luminosity spanning a
narrow redshift range. It may also be interesting to work out the
correlation of peaks with a fixed height but identified at smoothing
radii $R>R_f$, which can be thought as mimicking the statistical
properties of haloes above a given mass. However, we will not consider
this correlation here since it requires a solution to the
cloud-in-cloud problem~\cite{Bondetal1991} at the location of  density
maxima.

Let $\npk=\npk(\nu)$ hereafter denote the differential density of
peaks in the range $\nu$ to $\nu+\dd\nu$. The expectation value of the
product of the local peak densities that appears in
eq.~(\ref{eq:2pk1}) is then
\begin{widetext}
\begin{equation}
\xpk\!\left(\nu_1,\nu_2,r\right)=
\frac{1}{\la\npk\ra^2}\frac{5^5 3^4}{\left(2\pi\right)^6}\,
R_\star^{-6}\left(1-\gamma^2\right)^{-1}\,\int\!\!\prod_{i=1,2}
\left\{\dd u_i\dd v_i\dd w_i\,F\!\left(u_i,v_i,w_i\right)\right\}
\Phi_0\!\left(\nu_1,\nu_2,u_1,u_2,r\right)\,e^{-\bar{Q}}\;,
\end{equation}
where 
\begin{eqnarray}
\lefteqn{\Phi_0\!\left(\nu_1,\nu_2,u_1,u_2,r\right)=
\left\{u_1 u_2\left[\psi(r)+\gamma^2\xi(r)\right]\right.} && \\ &&
+\left. \nu_1\nu_2\left[\xi(r)+\gamma^2\psi(r)\right]-2\gamma^2
\left(u_1 u_2+\nu_1\nu_2\right)\Sigma(r)-\gamma\left(u_1\nu_2+u_2\nu_1
\right)\left[\xi(r)+\psi(r)-\left(1+\gamma^2\right)\Sigma(r)\right]
\right\}\,\left(1-\gamma^2\right)^{-2} \nonumber\;, 
\end{eqnarray}
\end{widetext}
is equation~(\ref{eq:average2}) averaged over the relative orientation of 
the frames spanned by the eigenvectors of $\zeta_1$ and $\zeta_2$.
$\Phi_0$ depends on the separation $r$ through the correlation functions 
$\xi(r)$, $\Sigma(r)$ and $\psi(r)$ only. Furthermore, the quadratic form 
$\bar{Q}$ simply is
\begin{equation}
2\bar{Q}=\nu_1^2+\frac{\left(u_1-\gamma\nu_1\right)^2}{1-\gamma^2}
+15 v_1^2+5 w_1^2 + 1\leftrightarrow 2
\label{eq:qform2}
\end{equation}
in the variables~(\ref{eq:newset}).

The integration over the variables $v_i$ and $w_i$ is lengthy but
straightforward. We refer the reader to BBKS for the details since the 
calculation now proceeds along similar lines. Let us mention that the 
allowed domain of integration is the interior of a triangle bounded by 
the points $(0,0)$, $(u_i/4,-u_i/4)$ and $(u_i/2,u_i/2)$. 
As shown in BBKS, the differential density of peak
of height $\nu$ can be cast into the form
\begin{equation}
\npk(\nu)=\frac{1}{\left(2\pi\right)^2 R_\star^3}\,e^{-\nu^2/2}\,
G_0\!\left(\gamma,\gamma\nu\right)\;,
\label{eq:npk}
\end{equation}
where $G_0$ is the zeroth moment of the peak curvature $u$. Higher moments 
are written in explicit compact form as
\begin{equation}
G_n\!\left(\gamma,\omega\right)=\int_0^\infty\!\!\dd x\,x^n f(x)
\frac{e^{-(x-\omega)^2/2(1-\gamma^2)}}{\sqrt{2\pi\left(1-\gamma^2\right)}}\;.
\label{eq:gk}
\end{equation}
Using this result, the correlation of peaks can be rearranged as follows~:
\begin{widetext}
\begin{equation}
\xpk\!\left(\nu_1,\nu_2,r\right)=G_0\!\left(\gamma,\gamma\nu_1\right)^{-1}
G_0\!\left(\gamma,\gamma\nu_2\right)^{-1}\,\int\!\!\prod_{i=1,2}
\left\{\dd u_i\,f(u_i)\frac{e^{-(u_i-\gamma\nu_i)^2/2(1-\gamma^2)}}
{\sqrt{2\pi\left(1-\gamma^2\right)}}\right\}\,
\Phi_0\!\left(\nu_1,\nu_2,u_1,u_2,r\right)\;.
\end{equation}
\end{widetext}
For sake of completeness,
\begin{eqnarray}
\lefteqn{f(x)=\frac{1}{2}\left(x^3-3x\right)
\left\{{\rm Erf}\left[\sqrt{\frac{5}{2}}x\right]
+{\rm Erf}\left[\sqrt{\frac{5}{2}}\frac{x}{2}\right]\right\}} \\ && 
+\sqrt{\frac{2}{5\pi}}\left[\left(\frac{31x^2}{4}+\frac{8}{5}\right)
e^{-5x^2/8}+\left(\frac{x^2}{2}-\frac{8}{5}\right)e^{-5x^2/2}\right]
\nonumber
\label{eq:fx}
\end{eqnarray}
as demonstrated in BBKS, who noted also that the asymptotic limits of
this function include a cancellation to eighth order at small $x$, and
the $x^3$  law expected for density maxima at large $x$.

The integration over $x$ must  generally be done numerically. It is
worth noticing that, while the exponential
$\exp[-(x-\omega)^2/2(1-\gamma^2)]$ decays rapidly to zero, $x^n f(x)$
are monotonically and rapidly rising. As a result, the functions
$G_n(\gamma,w)$ are sharply peaked around their maximum. For large
values of $\omega$, we find that $G_0$ and $G_1$ asymptote to
\begin{eqnarray}
G_0(\gamma,\omega) &\approx& \omega^3-3\gamma^2\omega+B_0(\gamma)\,
\omega^2\, e^{-A(\gamma)\omega^2} \\
G_1(\gamma,\omega) &\approx& \omega^4+3\omega^2\left(1-2\gamma^2\right)
+B_1(\gamma)\,\omega^3\, e^{-A(\gamma)\omega^2} \nonumber \;.
\label{eq:gasymptote}
\end{eqnarray}
The coefficients $A(\gamma)$, $B_0(\gamma)$ and $B_1(\gamma)$ are 
obtained from the asymptotic expansion of the Error function that
appears in eq.~(\ref{eq:fx}). We have explicitly
\begin{equation}
A=\frac{5/2}{\left(9-5\gamma^2\right)},~
B_0=\frac{432}{\sqrt{10\pi}\left(9-5\gamma^2\right)^{5/2}},~
B_1=\frac{4 B_0}{\left(9-5\gamma^2\right)} \;.
\end{equation}
The rest of the calculation is easily accomplished. The 2-point
correlation function of peaks eventually reads
\begin{widetext}
\begin{eqnarray}
\lefteqn{\xpk(\nu_1,\nu_2,r)=\left\{\left(\nu_1-\gamma\mcur_1\right)
\left(\nu_2-\gamma\mcur_2\right)\,\xi(r)\right.}
&& \\ && \left. +\left(\mcur_1-\gamma\nu_1\right)\left(\mcur_2-\gamma
\nu_2\right)\,\psi(r)-\left[\left(\nu_1-\gamma\mcur_1\right)\left(\gamma
\nu_2-\mcur_2\right)+\left(\gamma\nu_1-\mcur_1\right)\left(\nu_2-\gamma
\mcur_2\right)\right]\,\gamma\Sigma(r)\right\}
\left(1-\gamma^2\right)^{-2} \nonumber \;,
\end{eqnarray}
\end{widetext}
where we have introduced the mean curvature 
$\mcur(\gamma,\gamma\nu)=G_1/G_0$. Also, the notation is such that
$\mcur_i=\mcur(\gamma,\gamma\nu_i)$. 
The function $\mcur(\gamma,\gamma\nu)$ is accurately fitted by eq.~(4.4)
of BBKS, which is constructed to match the asymptotic 
large $\nu$ expansions of $G_0$  and $G_1$ given in eq.~(\ref{eq:gasymptote}). 
In the special case $\nu_1=\nu_2=\nu$, the 2-point correlation of 
peaks simplifies to
\begin{equation}
\xpk\!\left(\nu,r\right)=b_\nu^2\!\left(\nu,\gamma\right)\,\xi(r)
+b_\eta\!\left(\nu,\gamma\right)\,\Sigma(r)
+b_\zeta^2\!\left(\nu,\gamma\right)\,\psi(r)\;,
\label{eq:main}
\end{equation}
where the bias functions $b_\nu$, $b_\eta$ and $b_\zeta$ are
\begin{eqnarray}
b_\nu\!\left(\nu,\gamma\right) &=& \frac{\nu-\gamma\mcur}{1-\gamma^2} 
\nonumber \\
b_\zeta\!\left(\nu,\gamma\right) &=& \frac{\mcur-\gamma\nu}{1-\gamma^2}
\nonumber \\
b_\eta\!\left(\nu,\gamma\right) &=& 2\,\gamma\, b_\nu(\nu,\gamma)\,
b_\zeta(\nu,\gamma)\;.
\label{eq:biases}
\end{eqnarray}
The sign convention is chosen such that all three bias parameters are
positive when $\nu\rightarrow\infty$. Notice that $b_\nu$ is precisely
the amplification factor found by BBKS when derivatives of the density
correlation function are neglected.

Equation~(\ref{eq:main}), which holds for any value of the peak height
$\nu$  and the smoothing length $R_f$, is the main result of this
Section.  It describes  the asymptotic behaviour of the peak
correlation function in the limit where the correlation functions 
$\xi(r)$, $\Sigma(r)$ and $\psi(r)$ are much less than unity. 

\begin{figure}
\center\resizebox{0.45\textwidth}{!}{\includegraphics{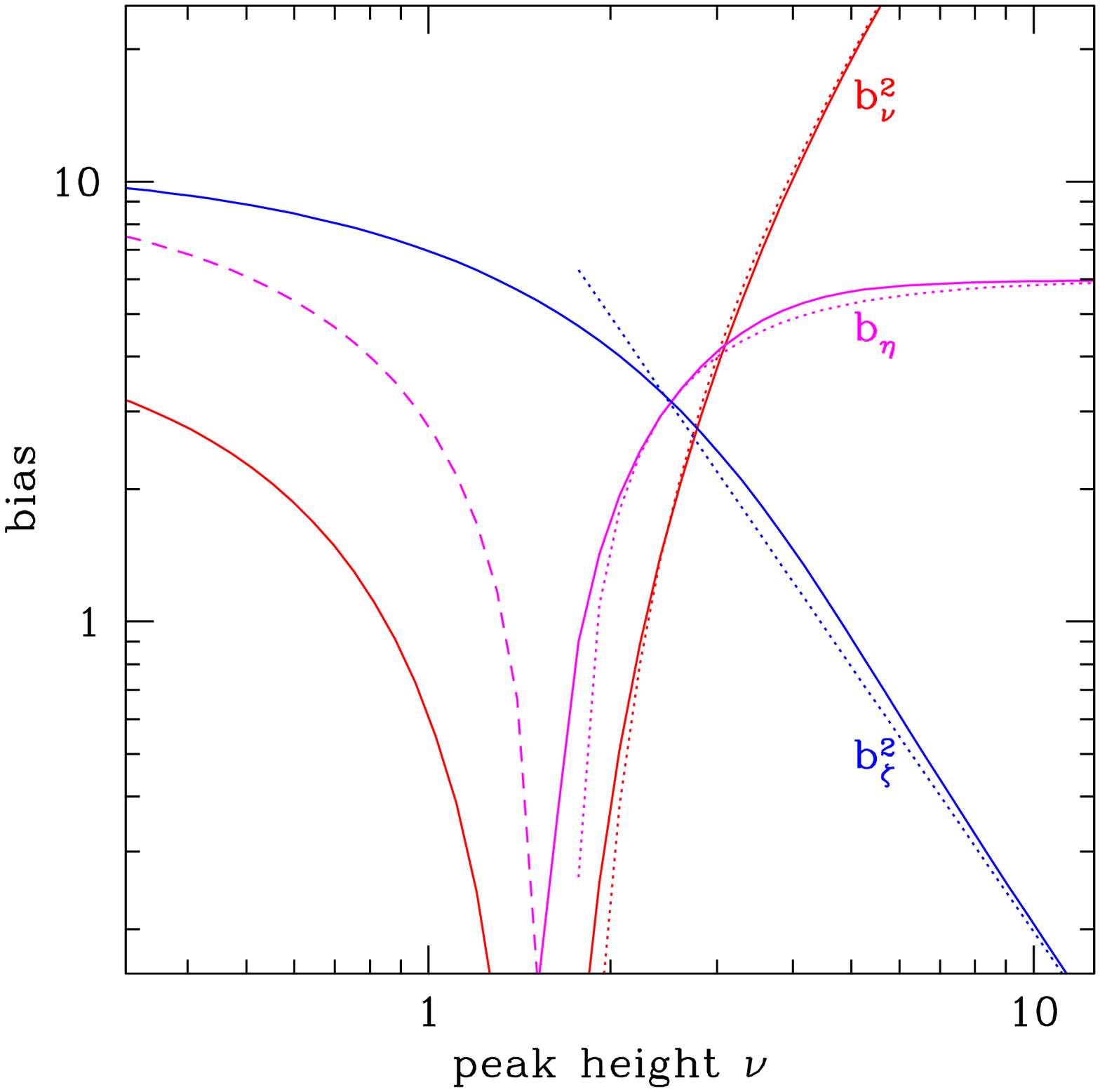}}
\caption{Bias factors $b_\nu^2(\nu,\gamma)$, $b_\eta(\nu,\gamma)$ and
$b_\zeta^2(\nu,\gamma)$ as a function of the peak height $\nu$. The
density field is smoothed on scale $R_f=5\hmpc$ with a Gaussian
filter. This leads to a correlation strength $\gamma=0.676$.  Dashed
curves indicate negative values. The dotted curves are the asymptotic
expansions given in eq.~(\ref{eq:basymptote}).}
\label{fig:biases}
\end{figure}

\subsection{The bias parameters}
\label{sub:biases}

To gain some insight into the behaviour of the peak correlation
function  $\xpk(\nu,r)$, we have plotted in Fig.~\ref{fig:biases} the
biasing parameters  $b_\nu$, $b_\eta$ and $b_\zeta$ as a function of
the peak height. Again, the  density field is smoothed  on scale
$R_f=5\hmpc$ with a Gaussian filter. The dotted curves show the
following large $\nu$ approximations,
\begin{eqnarray}
b_\nu(\nu,\gamma) &\approx& \nu-\frac{3}{\nu} \nonumber \\
b_\zeta(\nu,\gamma) &\approx& \frac{3}{\gamma\nu} \nonumber \\
b_\eta(\nu,\gamma) &\approx& 6\left(1-\frac{3}{\nu^2}\right)\;, 
\label{eq:basymptote}
\end{eqnarray} 
obtained from the asymptotic expansions of $G_0$ and $G_1$
(eq.~\ref{eq:gasymptote}). They provide a good match to the bias
parameters when the peak height is larger than $\simeq 2$. As we can
see, $b_\eta$ tends towards the constant value of 6 when
$\nu\rightarrow\infty$. Moreover, for a threshold height less than
unity, $b_\eta$ is negative and of absolute magnitude larger 
than $b_\nu^2$.  This is also true in the intermediate region  $\nu\sim
1-2$. For these threshold heights, both $b_\nu$  and $b_\eta$ vanish
while the bias parameter  $b_\zeta$ is of the order of a
few. Consequently, the correlation of density maxima, albeit weak for
peak heights of the order of unity, never cancels out. Overall,
retaining the density correlation  $\xi(r)$ solely is not a reasonable
approximation when the peak height  does not exceed $\nu\lsim
4$. Although the exact value of the bias parameters changes somewhat
with the smoothing scale $R_f$, their global behaviour varies little as
$\gamma$ weakly depends on the filtering scale. Therefore, the above
statements hold regardless of the exact amount of smoothing.

\subsection{Peak biasing~: nonlinear and local ?}

Equation~(\ref{eq:main}) clearly differs from the linear, local relation
$\xpk(\nu,r)=b_\nu^2\xi(r)$ that would be expected if the peak
overdensity $\delta\npk(\vx)=\npk(\vx)/\la\npk\ra$ were related to the
underlying density field through the linear mapping $\delta n_{\rm
pk}(\vx)=b_\nu\,\nu(\vx)$. However, it is worth noticing that
eq.~(\ref{eq:main}) is compatible with a nonlinear, local,
deterministic biasing  relation involving a differential
operator. Namely, it can be explicitly  checked that
\begin{equation} 
\delta\npk(\vx)=b_\nu\nu(\vx)+b_\zeta u(\vx)\;,
\label{eq:nllocbias}
\end{equation}  
where $u(\vx)=-\nabla^2\delta(\vx)/\sigma_2$, leads to the correlation
function ~(\ref{eq:main}). This demonstrates that, at large distances,
$\xpk(\nu,r)$ can be thought as arising from a specific case of
nonlinear local bias. We will see later (Sec.~\ref{sec:pkstream}) that
this local mapping is also consistent with the peak pairwise velocity
at first order.

To make connection with the formalism introduced 
by~\cite{FryGaztanaga1993}, we may conceive of a Taylor series
\begin{equation}
\delta\npk=\sum_{i=0}^\infty\frac{b_\nu^{(i)}}{i!}\nu^i
+\sum_{i=0}^\infty\frac{b_\zeta^{(i)}}{i!}u^i+\dots
\end{equation} 
to describe the properties of the peak distribution at all separations
and filtering scales. An  expansion of the peak correlation
$\xpk(\nu,r)$ beyond leading order will be required to determine the
values of the $b_\nu^{(i)}$ and $b_\zeta^{(i)}$ when $i>1$. Higher
derivatives of the density field may also contribute to this general
expression. However, (nonlocal) integrals of the linear density
correlation are  expected only in the evolved matter distribution when
the  non-Gaussianity induced by gravitational clustering is
significant.

Finally, it is worth noticing that, upon Fourier transformation, the 
peak power spectrum reads
\begin{equation}
P_{\rm pk}(\nu,k)=b_{\rm pk}^2\left[1+\frac{\sigma_0^2b_\eta}
{\sigma_1^2b_\nu^2}\,k^2+\frac{\sigma_0^2b_\zeta^2}{\sigma_2^2 b_\nu^2}
\, k^4\right]P(k)\;,
\label{eq:pspk}
\end{equation}
where $b_{\rm pk}\equiv b_\nu/\sigma_0$ and $P(k)$ is the power
spectrum  of the smoothed density field. The exact amount of
scale-dependence  induced by the nonlinear bias depends upon the exact
value of $\nu$  and $R_f$. We defer a thorough investigation of this
effect to a future work.

\section{Clustering of density peaks in Gaussian Initial Conditions}
\label{sec:pkclust}

After a brief discussion on the  peak-background split, we focus on
the acoustic signature in the 2-point correlation of density
maxima. We examine how the baryon acoustic oscillation changes with
the filtering, the  threshold height $\nu$ and the small-scale
behaviour of the transfer function. We find that the extra
contributions $b_\eta\Sigma(r)$ and $b_\zeta^2\psi(r)$ to the linear
relation $\xpk(\nu,r)=b_\nu^2\xi(r)$  can boost significantly the
contrast of the acoustic peak.

\subsection{Filtering scale and peak height}

The peak height $\nu$ and the filtering radius $R_f$ could in
principle be treated as two independent variables.  However, in order
to make as much connection with dark matter haloes (and, to a lesser
extent, galaxies) as possible,  we will follow the Press-Schechter
prescription~\cite{PressSchechter1974}   which is based on the
critical density criterion issued from the  spherical collapse
dynamics~\cite{GunnGott1972}.  Namely, we assume that density maxima
with peak height  $\nu=\dsc(z)/\sigma_0(R_f)$ identified in the
primeval density field smoothed at scale $R_f$ are related to dark
matter haloes of mass $M_f$ collapsing at redshift $z$. Moreover,  we
will only present results at redshift $z=0$, at which the linear
critical density for (spherical) collapse is $\dsc=1.673$, and the
characteristic  mass for clustering is $M_\star\approx 3.5\times
10^{12}\mdh$. While there is a direct correspondence between  the
massive  cluster-sized  haloes in the evolved density field and the
largest maxima  of the initial density field, it is unclear the extent
to which galaxy-sized haloes trace the initial density maxima
~\cite{galaxypeak}. For this reason, we will only consider mass scales
in the range $M_f\bsim M_\star(0)$ or, equivalently, a smoothing
radius $R_f\sim 1\hmpc$.

We note that the spherical infall model provides a local approximation
to the collapse of a perturbation. However, in the Press-Schechter
approach, it is applied to random points in space and leads to linear
local biasing at large scales~\cite{MoWhite1996,ShethTormen1999}
while, in the present work, it is applied to density maxima and leads
to the specific type of nonlinear local biasing exemplified by
eq.~(\ref{eq:nllocbias}).

\subsection{Peak-background split and the halo multiplicity function}

Before illustrating the impact of derivatives of the density field on
the baryon acoustic signature, we note that, in the limit $\nu\gg 1$,
the peak correlation is amplified by an effective bias $b_\nu^2$ which
is significantly smaller than the value $\nu^2/\dsc$ derived for
thresholded regions~\cite{Kaiser1984}. As recognised in BBKS,  this
difference arises from the correlation between the peak height $\nu$
and the peak curvature $u$.  More precisely, in the spherical infall
model, the linear Lagrangian bias $b_{\rm pk}^2=\xpk/(\sigma_0^2\xi)$
of high density peaks that are collapsing at redshift $z$ evaluates to
\begin{equation} 
b_{\rm pk}\approx \frac{\nu^2-3}{\dsc}
\label{eq:bpk}
\end{equation} 
in the limit $r\gg 1$. This should be compared to the expression derived  
in~\cite{MoWhite1996,ColeKaiser1989} from the Press-Schechter formalism
~\cite{PressSchechter1974,Bondetal1991},
\begin{equation} 
b_{\rm MW}=\frac{\nu^2-1}{\dsc}\;,
\label{eq:bmw96}
\end{equation} 
In this second approach, the clustering of haloes is described by the
properties  of regions above a given density threshold. In both cases
however, the Kaiser limit $\nu^2/\dsc$ is recovered. This, however,
does not apply to the bias factor derived by~\cite{ShethTormen1999}
using the ellipsoidal collapse,
\begin{equation}
b_{\rm ST}\approx\frac{a\nu^2-1}{\dsc}\;,
\label{eq:bst99}
\end{equation}  
where $a\simeq 0.7$. Assuming the peak-background split holds
~\cite{Kaiser1984}, these various bias parameters predict multiplicity
functions $\nu f(\nu)$~\cite{Bondetal1991} that have quite a different
behaviour in the limit of large threshold heights.  In particular, the
Sheth-Tormen (ST) multiplicity function is proportional to
$\nu\exp(-a\nu^2/2)$~\cite{ShethTormen2002}, and exponentially
deviates from the scaling inferred from $b_{\rm pk}$ and $b_{\rm MW}$,
which is $\nu f(\nu)\propto\nu^3\exp(-\nu^2/2)$ and
$\propto\nu\exp(-\nu^2/2)$, respectively. It is worth emphasising that
the factor $a=0.707$ was essentially determined by the  number of
massive haloes in the GIF simulations~\cite{Kauffmannetal1999}  and,
therefore, is not a direct outcome of the ellipsoidal collapse
dynamics. In fact, there is no compelling theoretical reason for a halo
mass function whose high-mass end deviates exponentially from the
scaling $\exp(-\nu^2/2)$. Furthermore, recent lines of evidence
suggest that the high-mass tail, while being above the Press-Schechter
(PS) mass function ~\cite{PressSchechter1974}, may depart from the
Sheth-Tormen scaling ~\cite{highmassend}.

\begin{figure}
\center\resizebox{0.45\textwidth}{!}{\includegraphics{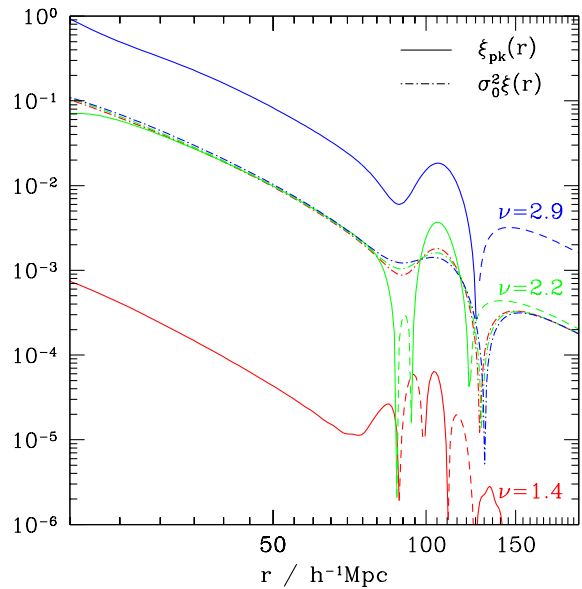}}
\caption{The peak correlation $\xpk(\nu,r)$ (solid curves) for three
different smoothing lengths $R_f=2$, 4 and $6\hmpc$ (from bottom to
top).  These correspond to a mass scale $M_f=9.5\times 10^{12}$,
$7.6\times 10^{13}$ and $2.6\times 10^{14}\mdh$, respectively.   A
peak height $\nu=\dsc/\sigma_0$ is adopted and yields the values
$\nu=1.40$, 2.15 and 2.88, respectively. The density correlation
$\sigma_0^2\xi(r)$ is plotted as the dotted-dashed curve. Dashing
indicates negative values.  The acoustic signature in the peak
correlation depends on the threshold height $\nu$ through the bias
parameters   $b_\nu$, $b_\eta$ and $b_\zeta$. Results are shown for
the $\Lambda$CDM cosmology.}
\label{fig:xpk1}
\end{figure}

In our opinion, it is likely that the true multiplicity function scales
as $\exp(-\nu^2/2)$ in the limit of large $\nu$. This would lead to a
different parametrisation of the halo bias and mass function. Given
the lack of a convincing physical  description of these quantities, 
one may, for instance,  consider a phenomenological bias of the form
\begin{equation}
b_{\rm L}=\frac{1}{\dsc}\left(\nu^2-c_1+\frac{c_2}{\nu^{2p}+c_3}\right)
\label{eq:lbias}
\end{equation}
which, for a peak-background split, leads to a multiplicity function
\begin{equation}
\nu f(\nu)\propto \left(1+\frac{c_3}{\nu^{2p}}\right)^{c_2/2pc_3}
\nu^{c_1}e^{-\nu^2/2}\;.
\label{eq:fnu}
\end{equation}
For $c_1\lsim 3$ and $c_3\sim c_2/c_1$ (which guarantees   $b_{\rm
L}\sim 0$  in the limit $\nu\rightarrow 0$), the
biasing~(\ref{eq:lbias}) closely follows   the peak scaling
eq.~(\ref{eq:bpk}) at large mass and,  simultaneously, exhibits an
upturn  at low mass. Unfortunately, such a bias cannot be derived from
an excursion set approach (upon which PS and ST are based), where 
$c_1=1$ invariably. This issue, which lies beyond the scope of the 
present paper, will be examined in a separate paper.

\subsection{Baryon acoustic signature}

We now turn to the behaviour of the peak correlation function.
$\xpk(\nu,r)$ is shown in Fig.~\ref{fig:xpk1} for a filtering length
$R_f=2$, 4 and $6\hmpc$. The mass enclosed in the Gaussian window thus
is $M_f=9.5\times 10^{12}$,  $7.6\times 10^{13}$ and $2.6\times
10^{14}\mdh$, respectively. To illustrate, we  have adopted a peak
height $\nu=\dsc(z=0)/\sigma_0$ such that $\nu=1.4$, 2.1 and 2.9,
respectively. In the spherical infall dynamics, a  top hat  overdensity
enclosing a similar amount of mass would collapse at redshift  $z\sim
0$. Furthermore, the density correlation  $\sigma_0^2\xi(r)$ is also
shown for comparison as the  dotted curve.

The three correlations considered here exhibit a very different
behaviour that reflects the strong dependence of the bias factors
$b_\nu$, $b_\eta$ and $b_\zeta$ on the threshold height (see
Sec.~\ref{sub:biases}). In particular,  we find $b_\nu=-0.057$, 0.847
and 1.771 with increasing smoothing radius. As a result, for
$R_f=2\hmpc$, the contribution of the term $b_\zeta^2\psi(r)$
dominates the others and strongly suppresses the amplitude of
$\xpk(\nu,r)$ relative to that of the density correlation.  This term
has the sign of $\psi(r)$ and features several oscillations across the
BAO scale (see Figures~\ref{fig:xir1} and \ref{fig:xir2}).  However,
for peaks of threshold height $\nu=2.1$ identified at smoothing scale
$R_f=4\hmpc$, $b_\zeta^2\psi(r)$ merely contributes to decrease the
level of the minimum at distance $r\sim 90-95\hmpc$.  Interestingly,
the term $b_\eta^2\Sigma(r)$ boosts significantly the contrast of the
acoustic peak. This effect is still present, albeit weaker, for
$\nu=2.9$ and $R_f=6\hmpc$. We also note that zero-crossings of
$\xpk(\nu,r)$ do not generally coincide with those of $\xi(r)$, in
agreement with numerical studies of the clustering of density
maxima~\cite{Coles1989, Lumsdenetal1989}.

\begin{figure}
\center\resizebox{0.45\textwidth}{!}{\includegraphics{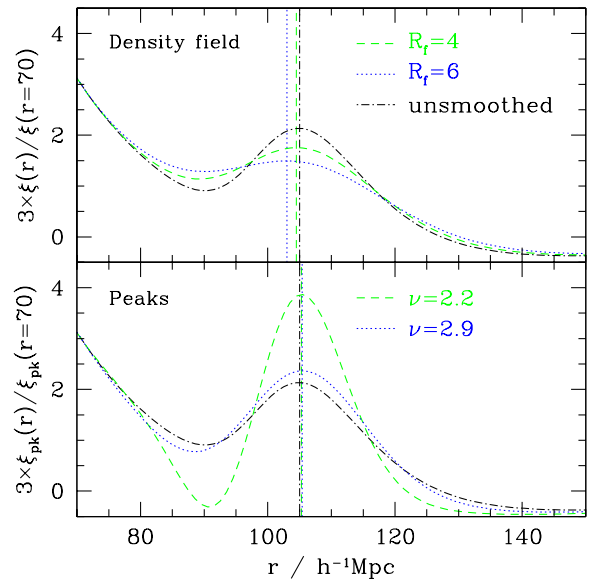}}
\caption{A comparison between the density correlation $\xi(r)$ (top
panel)  and the peak correlation $\xpk(\nu,r)$ (bottom panel) around
the BAO.   The density field is smoothed with a Gaussian filter of
width $R_f=4$ and $6\hmpc$.  The corresponding value of peak height is
$\nu=2.1$ and 2.9, respectively.  For clarity, all the correlations
have been rescaled such that, at  separation $r=70\hmpc$, their
amplitude is equal to 3. Also shown as the dotted-dashed line is the
(unsmoothed) linear matter correlation. The vertical dashed  lines
indicate the position of the local maximum.  The presence of
$b_\eta^2\Sigma(r)$ in the peak correlation restores, and even
amplifies the acoustic peak otherwise smeared out by  the large
filtering. $b_\eta^2\Sigma(r)$ also acts to reduce the  shift induced
by the smoothing. Results are shown for the $\Lambda$CDM cosmology.}
\label{fig:bao1}
\end{figure}

Fig.~\ref{fig:bao1} further illustrates the sharpening of the acoustic
peak due to  correlations among derivatives of the density field.  The
density and the peak correlations are compared in the neighbourhood  of
the acoustic  feature for the smoothing radii $R_f=4$ and $6\hmpc$
considered above.  To  emphasise the contrast of the acoustic peak,
all the correlations have been rescaled  such that, at a distance
$r=70\hmpc$, their amplitude is equal to 3.  Fig.~\ref{fig:bao1}
nicely demonstrates the large impact of $b_\eta^2\Sigma$, which fully
restores the acoustic signature of $\xpk(\nu,r)$ otherwise  smeared
out by the large filtering.  The contrast of the acoustic peak can
even be enhanced relative to that of the unsmoothed ($R_f=0.1\hmpc$)
linear density correlation (dotted-dashed line). The effect is
strongest for the density peaks identified at the  smaller smoothing,
$R_f=4\hmpc$. For  these maxima, the difference  between the height of
the (negative)  minimum at $r\simeq 90\hmpc$ and the maximum at
$\simeq 105\hmpc$ is twice as large as in the linear density
correlation. The enhancement is somewhat smaller, roughly 20 per cent,
for the peaks at the filtering scale  $R_f=6\hmpc$. This shows that
density maxima behave rather differently than linearly biased tracers
of the  density field, whose acoustic signature cannot be larger than
that  of the linear matter correlation~\cite{Eisensteinetal2007a}.

We now concentrate on the vertical lines which indicate the position
of the local maximum. On the one hand, the top panel shows that
smoothing in the density correlation  generates a shift towards
smaller scales, because the acoustic feature  is not quite symmetric
around its maximum ~\cite{Smithetal2007}. On the other hand, the
presence of $b_\eta^2\Sigma$ in the peak correlation acts in the
opposite sense and compensates for the shift induced by the
smoothing. We find the maximum to be close to  its linear value
$\approx 105.0\hmpc$ in both cases. More precisely, there is a small
shift of $\lsim 0.4$ per cent towards larger scales.

\subsection{Sensitivity to the filter shape and the transfer function}

As discussed in BBKS, the filtering of the density field is an
essential operation for power spectra covering a wide range of
wavenumbers. However, the optimal choice of filter is disputable.
Furthermore, the (analytic) properties of the filtered density field
can depend significantly upon the amount of power in small-scale
fluctuations. It is, therefore, important to assess the influence of 
the smoothing operation and the small-scale transfer function on the
baryon acoustic signature in the correlation of density peaks.

To this purpose, we have repeated the numerical calculation of
$\xpk(\nu,r)$ using  a top hat filter.  To avoid divergence of the
spectral moments and the correlation functions, we have introduced a
high-$k$ cutoff whose functional form is motivated by the damping of
fluctuations due to the free streaming of  the dark matter
particle(s). So far, we have considered a CDM cosmology in which the
velocity dispersion of the dark matter particle is negligible. By
contrast, in Warm Dark Matter (WDM)  cosmologies, the dark matter
candidate(s) can suppress the matter power spectrum on  galaxy scales
$r\sim 0.1\hmpc$~\cite{Bondetal1980}. The latter can be approximated
as $P_{\rm WDM}(k)=T^2(k) P_{\rm CDM}(k)$, where the transfer function
that accounts for the free-streaming cutoff has the form~\cite{tkwdm}
\begin{equation}
T(k)=\left[1+\left(\alpha k\right)^{2 p}\right]^{-5/p}\;.
\label{eq:transfer}
\end{equation}
Here, $p\approx 1.12$ and $\alpha$ depends upon the properties of
the dark matter particles. Typically, $0.01\lsim\alpha\lsim 0.1$ for
thermal relics of mass $\sim 1-10\keV$. 

In spite of its compactness, the top hat filter has some inconvenient.
Firstly, because of its slowly decaying tail, it produces a density
field that is not differentiable for generic CDM power
spectra~\cite{Bondetal1991}. Secondly, it is good at discriminating
peaks from the background field so long as the height of the latter is
small, namely, when the background field is uncorrelated over scales
comparable to the filtering length. By contrast, the Gaussian filter
is less sensitive to high frequencies and thus fares better at picking
up smoother objects. Indeed, the ``true'' filter may lie between these
two extremes~\cite{Dalaletal2008}.  Notice that the sharp $k$-space
window will not be considered here as it leads to undesirable
oscillations at all separations.

\begin{figure}
\center\resizebox{0.45\textwidth}{!}{\includegraphics{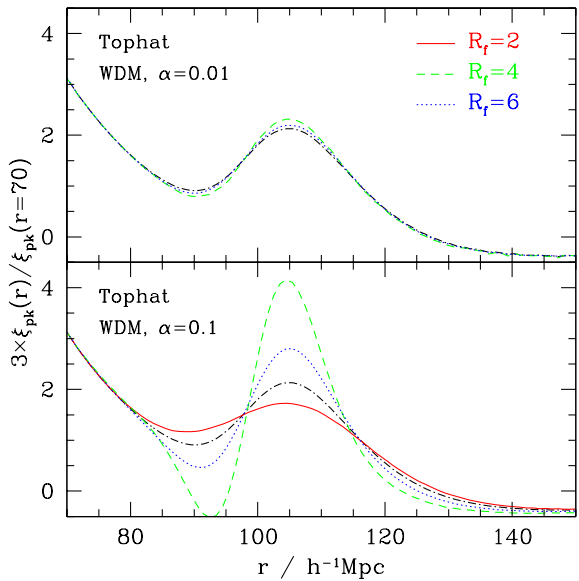}}
\caption{The correlation of density maxima that trace the density field
smoothed with a Top Hat filter. The smoothing radii $R_f=2$, 4 and
$6\hmpc$ correspond to a mass $M_f=2.5\times 10^{12}$, $2\times
10^{13}$ and $6.8\times 10^{13}\mdh$, respectively. Results are shown
for a WDM power spectrum with a cutoff scale $\alpha=0.01$ and
$0.1\hmpc$ (see text). The peak height is chosen such that
$\nu=\dsc/\sigma_0$, as before. In both panels, the dotted-dashed curve 
is the linear matter correlation.  The peak correlation $\xpk(\nu,r)$ 
for $\alpha=0.01$ and $R_f=2$ is not shown as it is too much affected 
by numerical noise.}
\label{fig:bao2}
\end{figure}

Fig.~\ref{fig:bao2} shows the baryon acoustic peak in the correlation
of density maxima for the smoothing radii used in Fig.~\ref{fig:xpk1}
and~\ref{fig:bao1}. Note, however, that the filter mass scale is now
roughly four times smaller than with the  Gaussian window. The peak
correlation is plotted for two values of  the free-streaming cutoff,
$\alpha=0.01$ and 0.1 (top and bottom panels).  Also shown in  both
panels for comparison is the linear matter correlation (dotted-dashed
line). At smoothing length $R_f=4$ and $6\hmpc$, the enhancement of
the acoustic peak is very significant for $\alpha=0.1$ while, for
$\alpha=0.01$, it is only 10-20 per cent. The main reason is a sharper
power spectrum, which leads to a larger contribution of the
correlations $\Sigma(r)$ and $\psi(r)$ to $\xpk(\nu,r)$. At
$R_f=4\hmpc$  for instance, the spectral width is $\gamma=0.48$ and
0.26 for  $\alpha=0.1$ and 0.01, respectively. This difference mostly
arises because of the second spectral moment, which increases from
$\sigma_2=0.40$ to 0.76 upon the decrease in the free-streaming
scale. Yet another interesting feature of Fig.~\ref{fig:bao2} is the
rather broad acoustic peak at filtering scale $R_f=2\hmpc$ (see bottom
panel), for which $\nu=0.96$.  This broadening follows from  the fact
that $b_\eta$ is negative at that value of threshold height. As a
consequence, the oscillatory pattern of  $\Sigma(r)$  across the BAO
(see Figures~\ref{fig:xir1} and \ref{fig:xir2}) smears out the
acoustic feature in $\xpk(\nu,r)$. As seen from Fig.~\ref{fig:biases},
this damping always  occurs at sufficiently low values  of the
threshold height, $\nu\lsim 1$ regardless of the filtering length. It
should also be noted that,  unlike the correlation of density maxima,
the BAO in the smoothed linear matter  correlation  is weakly
insensitive  to the small-scale behaviour  of the power spectrum. In
$\xpk(\nu,r)$ however, the BAO acquires an extra dependence upon the
high-$k$ tail of the transfer function through the correlation
functions $\Sigma(r)$ and $\psi(r)$.

To summarise, 
\begin{itemize}
\item{Both $\Sigma(r)$ and $\psi(r)$ contribute to the correlation
of density maxima and can affect the shape of the baryon acoustic 
signature for peak heights $\nu\lsim 4$.}
\item{$\psi(r)$ makes a significant contribution only in the range
$1\lsim\nu\lsim 2$, where $b_\nu$ and $b_\eta$ are much less than 
unity.}
\item{The contribution of $\Sigma(r)$ increases with the spectral
width $\gamma$. At constant filtering length, it increases with 
the amount of power suppression due to the small scale free 
streaming.}
\item{$b_\eta$ is positive (negative) for $\nu\bsim 1$ ($\nu\lsim 1$).
As a result, the baryon acoustic peak is generally enhanced in 
$\xpk(\nu,r)$ when $\nu\bsim 1$, and damped out when $\nu\lsim 1$. }
\end{itemize}
These results depend upon the exact shape of the filter and the
transfer function. Clearly however, the effect cannot be reduced  to a
simple rescaling of the linear matter correlation. This is due  to the
peculiar type of nonlinear local biasing, eq.~(\ref{eq:nllocbias}),
which involves the Laplacian of the density field.

\section{Pairwise velocity of density maxima}
\label{sec:pkstream}

Thus far, we have explored the BAO signature in the correlation of
maxima of the primordial density field.  However, pairwise motions
caused by small- and large-scale structures, redshift space
distortions etc. are likely to degrade the acoustic signature, leading
to a  broadening and, possibly, a shift of the acoustic peak.  A
thorough investigation of these effects  is postponed to a subsequent
work. Here, we consider a simple  model in  which the peak centres
evolve according to the Zeldovich ansatz. This allows us to calculate
the peak pairwise velocity at leading order, which is the main result
of this Section. We show that, at first order, the peak mean streaming
is consistent with the nonlinear local bias found in
Sec.~\ref{sec:pkclust}. Dynamical evolution is also briefly addressed
using  the pair conservation equation.

\subsection{Zeldovich approximation}

The Eulerian comoving position and proper velocity of a density peak 
can generally be expressed as a mapping
\begin{equation}
\vxpk=\vq+\vvs(\vq,a)\;,~~~\vvpk=a\,\dot{\vvs}(\vq,a)\;,
\label{eq:za} 
\end{equation}
where $\vq$ is the initial position, $\vvs(\vq,a)$ is the displacement
field and $a$ is the scale factor. A dot denotes a time derivative.
At first order, the peak position is described by the Zeldovich
approximation~\cite{Zeldovich1970}, in which the displacement
factorises  into a time and a spatial component,
\begin{equation}
\vvs=-D(a)\grad\Phi(\vq),~~~\dot{\vvs}=-\beta(a)\grad\Phi(\vq)\;.
\end{equation} 
Here, $\Phi(\vq)$ is the perturbation potential linearly extrapolated
to present time. Explicitly, $\Phi(\vq)=\phi(\vq,a)/4\pi G\rb(a)a^2 D(a)$
where $\phi(\vq,a)$ is the Newtonian gravitational potential,
$\rb(a)$ is the average matter density and $D(a)$ is the growth
factor. $\beta(a)=HDf$ is proportional to the logarithmic derivative
$f=\dd\ln D/\dd\ln a$, which scales as $f(a)\approx\Omega_m(a)^{0.6}$
for a wide range of CDM cosmologies ~\cite{Peebles1980}. Finally, 
$H(a)$ is the Hubble constant.

Such a simple model cannot account (among other things) for the
internal properties of peaks~\cite{GrinsteinWise1987}. Furthermore, it
provides a very limited description of the  late-time distribution of
density maxima such as cluster- or galaxy-size haloes
~\cite{zapeak,Coles1993}. Notwithstanding this, it is not intended to
be realistic, but only to capture the weakly nonlinear regime
reasonably well. A more sophisticated approach can be found
in~\cite{peakpatch} for instance.

The peak pairwise velocity, or mean
streaming ~\cite{Peebles1976,DavisPeebles1977}, is now obtained from
the statistics  of the (proper) matter velocity field
$\vv=-a\beta(a)\grad\Phi(\vq)$. The complication arises from the fact
that the latter has to be evaluated at those maxima of the density
field.

\begin{figure}
\center\resizebox{0.45\textwidth}{!}{\includegraphics{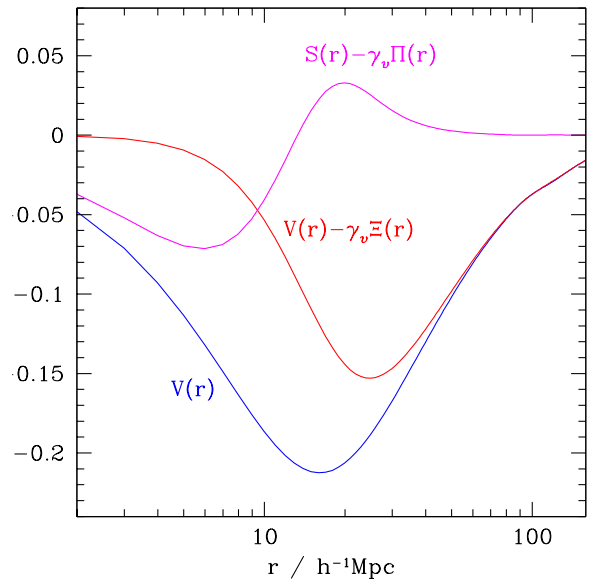}}
\caption{The correlations that contribute to the leading order mean
streaming of peak pairs, equation~(\ref{eq:pkstream}). These are
compared to the line of sight pairwise velocity ${\cal V}(r)$ of
ambient field points. Results are shown at a filtering scale
$R_f=5\hmpc$. The correlation ${\cal V}-\gamu\Xi$ is strongly damped
on scales less than the characteristic inter-peak distance  $\propto
R_f$ but, at large distances, it is unaffected by small-scale 
exclusion effects and closely follows the (scaled) mean streaming
of random field points. The correlation ${\cal S}-\gamu\Pi$ can 
significantly contribute to small-scale streaming motions when the 
peak height is $\nu\lsim 3$ (so that $b_\zeta\bsim b_\nu$). }
\label{fig:v12}
\end{figure}

\subsection{Mean streaming of peak pairs}

We introduce the normalised velocity field
$\vu(\vq)=\vv(\vq)/(a\beta(a)\sigma_{-1})$ for subsequent use,  and
define $u_{12}(r)$ as the average number weighted pairwise velocity
$[\vu(\vx_2)-\vu(\vx_1)]\cdot\rvh$ along the line of sight.

The calculation of the peak pairwise velocity is more intricate than
the peak correlation since we have three additional degrees of
freedom.  Nevertheless, it closely follows the analysis described in
Sec.~\ref{sec:pkcorr}.  Details of the calculation can be seen at
Appendix ~\ref{sec:pkstreamdetails}. The peak mean streaming weighted
by the  number density of peaks at $\vq_1$ and $\vq_2$ eventually 
reads
\begin{eqnarray}
\left(1+\xpk\right)u_{12} &=& 
\left[b_\nu(\nu_1,\gamma)+b_\nu(\nu_2,\gamma)\right]
\left({\cal V}-\gamu\Xi\right) \\
&& +\left[b_\zeta(\nu_1,\gamma)+b_\zeta(\nu_2,\gamma)\right]
\left({\cal S}-\gamu\Pi\right) \nonumber \;.
\label{eq:pkstream}
\end{eqnarray}
For comparison, the pairwise velocity of the matter distribution is
$[1+\sigma_0^2\xi(r)]u_{12}(r)=2\sigma_0{\cal V}(r)$.  The first term
in the right-hand side of eq.~(\ref{eq:pkstream}) is similar to the
mean streaming derived for locally biased tracers of the density
field~\cite{Shethetal2001}. The second term arises because of the
particular nature of the bias of density maxima. Interestingly,
eq.~(\ref{eq:pkstream}) can again be thought of as arising from the
nonlinear local bias eq.~(\ref{eq:nllocbias}) if we choose the peak
velocity  field to be
\begin{equation}
{\bf u}_{\rm pk}(\vx)=\vu(\vx)-\gamu\ve(\vx)\;.
\end{equation}
In this continuous approach, the peak velocity field is still unbiased
with respect to the matter velocity field $\vu(\vx)$, but it receives
a contribution from the first derivative of the density,
$\eta(\vx)=\nabla\delta(\vx)/\sigma_1$, that is  proportional to
$\gamu$.

The sign and the strength of the peak pairs flow depend upon the 
detailed behaviour of the functions ${\cal V}-\gamu\Xi$ and ${\cal
S}-\gamu\Pi$. As seen in Fig.~\ref{fig:v12} where $R_f=5\hmpc$ for
illustration, the former is negative at all separations $r\lsim
200\hmpc$. By contrast, the latter is positive at distances larger
than a few smoothing radii but goes negative at smaller scales,
regardless of the exact value of $R_f$.  The mean streaming ${\cal
V}(r)$ of random field points is also shown in Fig.~\ref{fig:v12} for
comparison. Peak-peak exclusion leads to a deficit of pairs at
separation $r\lsim R_\star$ comparable to the filtering scale. It adds
to the smoothing and further damps the relative velocity of peaks out
to distances that are much larger than the typical extent $\sim R_f$
of density maxima. This is the reason why  the correlation ${\cal
V}-\gamu\Xi$ is strongly suppressed relative to ${\cal V}(r)$ when
$r\lsim 30\hmpc$. Still, the term proportional to ${\cal S}-\gamu\Pi$
is most  strongly negative at distances of the order of the smoothing
length and, therefore, could restore significantly the small-scale
mean streaming when the threshold height is less than $\nu\lsim 3$
(for which $b_\zeta\bsim b_\nu$).  At large enough separations $r\bsim
50\hmpc$ however,  the mean streaming of peak pairs is unaffected by
small-scale  exclusion effects and closely tracks the pairwise
velocity ${\cal V}(r)$ of ambient field points.

\begin{figure}
\center\resizebox{0.45\textwidth}{!}{\includegraphics{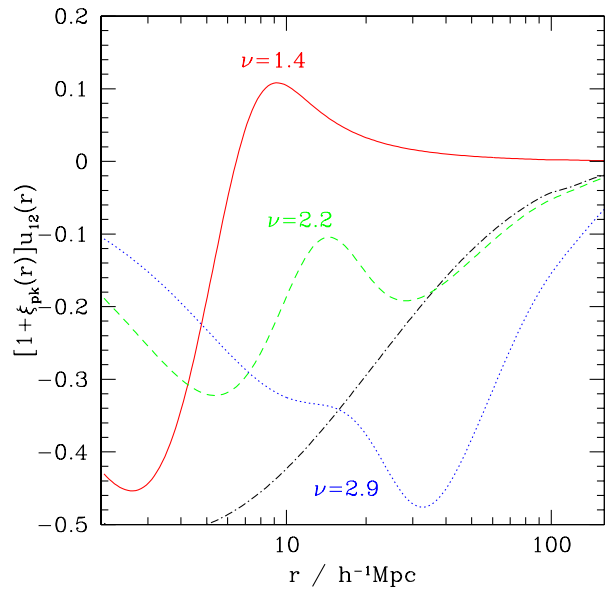}}
\caption{The mean streaming of peak pairs for density maxima 
identified at smoothing scale $R_f=2$, 4 and 6$\hmpc$ and with
peak height $\nu=1.40$, 2.15 and 2.88, respectively (as in 
Fig.~\ref{fig:xpk1}). The dotted-dashed curve shows the pairwise
velocity of the (unsmoothed) underlying density field. The 
smallest peaks tend to accrete onto high density maxima.  
However, they move apart from each other relative to the matter
distribution due to peak-peak exclusion.}
\label{fig:vpk1}
\end{figure}

To illustrate the impact of the correlation function ${\cal
S}-\gamu\Pi$ on the mean streaming, we show in Fig.~\ref{fig:vpk1} the
peak pairwise velocity for the smoothing radii $R_f=2$, 4 and 6$\hmpc$
considered in Sec.~\ref{sec:pkclust}  (recall that the peak height is
specified by the relation $\nu=1.673/\sigma_0(R_f)$). Also shown for
comparison is the mean streaming of the (unsmoothed) matter density
field (dotted-dashed curve).  At the smallest filtering scale for
which $\nu=1.4$, the bias parameters are $-b_\nu\approx 0.06\ll
b_\zeta\approx 2.32$ so that ${\cal S}-\gamu\Pi$ is the dominant
contribution at all separations. The resulting strong ``inward''
transport at distances less than a few $\hmpc$ reflects the fact that
these small peaks tend to accrete onto high density regions. At
separation $r\sim 10\hmpc$, there is a positive net flow presumably
owing to the fact that  the peaks fall onto nearby distinct overdense
regions. Notice that, at all separation, the mean streaming of these
maxima is larger than that of the density field, indicating that these
small peaks move apart from each other (in an average sense) relative
to the matter distribution.  At larger filtering scales, the
contribution of the first term in the right-hand side of
eq.(\ref{eq:pkstream}) increases with the smoothing length as seen
from the progressive disappearance of the broad bump at $r\sim
10\hmpc$. The maxima identified at $R_f=6\hmpc$ ($\nu=2.9$)  stream
towards each other relative to the underlying density field, but their
relative motion is strongly suppressed at distances $r\lsim 10\hmpc$
due to the exclusion effect mentioned above.

\subsection{Pair conservation equation}

So long as peaks do not merge, the time evolution of the peak
correlation function $\xpk$ is  governed by the pair conservation
equation. Transforming the time variable to the scale factor, this
equation can be written as
\begin{equation}
\frac{\partial\xpk}{\partial a}=-Df\,\sigma_{-1}\,\frac{1}{r^2}
\frac{\partial}{\partial r}\left[r^2\left(1+\xpk\right)u_{12}(r)
\right]\;,
\label{eq:pairs1}
\end{equation}
where $r$ and $u_{12}(r)$ are the comoving separation and scaled
pairwise velocity, respectively. The root-mean-square (rms) variance
$\sigma_{-1}$ (computed from eq.~\ref{eq:mspec}) defines the length
scale $\sigma_{-1}\approx 9.2\hmpc$  for the values of cosmological
parameters used here.

Following the approach outlined in ~\cite{Smithetal2008}, the general
solution of  eq.~(\ref{eq:pairs1}) can be found by solving the
characteristic  equation
\begin{equation}
\frac{\dd r}{\dd a}=D f\,\sigma_{-1}\,u_{12}(r)\;.
\label{eq:char}
\end{equation}  
This equation gives $r(a)$ which, upon insertion into the pair
conservation equation~(\ref{eq:pairs1}), allows us to write down  a
first order ordinary differential equation along the characteristics,
\begin{equation}
\frac{\dd\ln\left[1+\xpk(r,a)\right]}{\dd a}=-D f\,\sigma_{-1}\,
\frac{1}{r^2}\frac{\partial\!\left[r^2 u_{12}(r)\right]}{\partial r}\;,
\label{eq:char}
\end{equation}
where it is understood that $r=r(a)$~\cite{Smithetal2008}.  

We will not attempt to solve eq.~\ref{eq:char} since, as recognised
by~\cite{Smithetal2008,CrocceScoccimarro2008}, nonlinearity in the
divergence of the pairwise velocity, which is lacking here, is  a
crucial ingredient in the redshift evolution of the  baryon acoustic
signature.  Instead, we will simply estimate the first order change in
the  initial separation of peak pairs, $\Delta r_0$, induced by
coherent  motions  across the acoustic scale, $r_0\sim 105\hmpc$.  To
proceed, we assume that the peaks move according to the Zeldovich
ansatz described above, an approximation expected to be valid only in
the early (quasi-linear) stages of gravitational clustering. Owing to
the near constancy of the peak pairwise velocity at those scales, we
can write  $\Delta r_0\approx \sigma_{-1}u_{12}(r_0)\int D f\dd a$
where $\int D f\dd a\approx 0.56$.  For the maxima considered above,
we find $\Delta r_0(r_0=105)=+0.010$, -0.25 and -0.74$\hmpc$ with
increasing $R_f$. For comparison, $\Delta(r_0=105)\approx 0.21\hmpc$
for the dark matter. These values are consistent with those found
by~\cite{Smithetal2008}. Therefore, at the linear order, changes in
the acoustic signature of $\xpk(r)$ will be roughly at the percent
level.  This suggests that some of the enhancement of the BAO in the
initial correlation of density maxima may survive in the correlation
of high redshift density peaks. Clearly, a thorough numerical
investigation and detailed analytic modelling will be needed to
ascertain how much of this effect propagates into the late-time
clustering of galaxies.

\section{Conclusions}
\label{sec:conclusions}

We have investigated the strength of the baryon acoustic signature in
the 2-point correlation of maxima of the linear (Gaussian) density
field $\delta(\vx)$. To this purpose, we examined in
Sec.~\ref{sec:pkcorr} the large-scale asymptotics of the peak
correlation $\xpk(r)$ and derived  the leading order contribution,
eq.~(\ref{eq:main}).  In contrast to  the analysis of BBKS,  spatial
derivatives of the linear density correlation $\xi(r)$ were  included
in our derivation. These derivatives are not negligible for  generic
CDM power spectra, especially around the BAO scale where they  exhibit
large oscillations. we find that the leading asymptotic behaviour of
the peak correlation is governed by three terms~: a term previously
derived in BBKS plus  two terms involving the spatial derivatives
$\Sigma(r)$ and $\psi(r)$  of the linear density correlation. The
relative contribution of these functions is controlled by two
independent bias parameters, $b_\nu$  and $b_\zeta$ (The third being
$b_\eta=2\gamma b_\nu b_\zeta$, see  eq.~\ref{eq:biases}). We also
showed that the large-scale asymptotics of $\xpk(r)$ can be thought as
arising from a nonlinear, local biasing relation,
eq.~\ref{eq:nllocbias}, involving second derivatives of the density
field.

In Sec.~\ref{sec:pkclust}, we demonstrated that those extra terms 
have a large impact on the correlation of density maxima
in the vicinity of the BAO. The  results are sensitive to the exact
value of the threshold height $\nu$, the smoothing length $R_f$, the
filter shape and the high-$k$ tail of the transfer function. For the
Gaussian filter adopted throughout this paper, the contrast of the
baryon acoustic signature can be significantly enhanced relative to
that in the linear matter correlation when the peak height is in the
range $1\lsim\nu\lsim 3$. This boost originates from the oscillatory
behaviour  of $\Sigma(r)$ and $\psi(r)$ around the sound horizon
scale. For  instance, we find that, at filtering scale $M_f=8\times
10^{13}\mdh$,  the contrast of the BAO in the correlation of density
maxima of height is about twice as large as in $\xi(r)$. The
amplification fades as we  go to larger  peak height. For a peak
height of the order of  unity, $\xpk(r)$ can exhibit several bumps
which reflect those of $\psi(r)$ around the BAO scale. For a threshold
height less than $\lsim 1$,  the original acoustic peak is smeared out
by the negative contribution of the term $b_\eta\Sigma(r)$.

To avoid the divergence of the (fourth order) spatial derivative
$\psi(r)$ of the density correlation, we have filtered the density
field with a Gaussian window. The main drawback of this window
function is the lack of a well-defined mass and spatial extent
associated to the density fluctuations. A top hat filter appears better
motivated in the context of, e.g., the spherical infall model,
although it does not produce an infinitely differentiable  density
field for generic CDM power spectra. Furthermore, the
differentiability of the density field depends strongly upon the
small-scale behaviour of the transfer function. Fluctuations in the
matter density are damped on scales smaller than the free-streaming
length of the dark matter particle. In the CDM cosmology considered
here, the velocity dispersion of the dark matter particle is
negligible. By contrast, Warm Dark Matter (WDM)  particles such as
massive neutrinos for example can suppress the matter power spectrum
on galaxy scales~\cite{Bondetal1980}.  For these reasons, we also
discussed in Sec.~\ref{sec:pkclust} how the BAO changes with the
window function and the free-streaming cutoff. We found that the
correlation of density maxima is more sensitive to the properties of
the dark matter particle(s) than  the matter correlation itself. This
follows from the dependence of the peak biasing upon the second
derivatives of the density field. However, whether the baryon acoustic
signature in the clustering  of peaks varies  significantly with the
nature of dark matter remains  to be determined.

In Sec.~\ref{sec:pkstream}, we calculated the pairwise velocity of
peak pairs at leading order.  We showed that it is consistent with the
nonlinear local biasing relation inferred from the 2-point correlation
of density  maxima, provided that the peak velocity field receives  a
contribution from the gradient of the density field.  Explicitly, the
leading-order peak correlation and mean streaming  can be derived from
the nonlinear local biasing relation
\begin{eqnarray}
\delta\npk(\vx) &=& b_\nu\delta(\vx)-b_\zeta\nabla^2\delta(\vx)
\nonumber \\  \vu_{\rm pk}(\vx) &=& \vu(\vx)-\gamu\nabla\delta(\vx)\;,
\end{eqnarray} 
where we have dropped some factors for clarity and $\vu(\vx)$ is  the
linear matter velocity field.  This particular bias relation may be
helpful to translate the Lagrangian analysis performed in this paper
into quantitative predictions for the baryon oscillation in the low
redshift distribution of galaxies, which is currently the primary
observable proxy of the baryonic acoustic oscillations. Using the
formalism  introduced by ~\cite{FryGaztanaga1993}, one may conceive of
sophisticated extensions of the halo model~\cite{halomodel} that
would include derivatives of the density field, so as to ascertain how
much  of the amplification of the acoustic signature in the initial
clustering of density maxima propagates into the late-time correlation
of galaxies  and clusters.  Extensions which, as a general criterion,
reproduce the observed properties of the galaxy distribution, would
provide an interesting complement to current local biasing models.

We emphasise that the calculations presented in this paper are
performed in the initial conditions. As nonlinearities progress, the
late-time acoustic signature is smeared out by structure formation as
reported by many authors using N-body simulations~\cite{baosimu}.
This might explain why numerical investigations of the clustering of
dark matter haloes have not shown thus far any evidence for an
amplification of the BAO. Interestingly however, preliminary results
from  a large suite of N-body simulations hint at a an enhancement of
the contrast of the baryonic signature in the clustering  of low
redshift dark matter haloes~\cite{resmith}.  In this regard, it is
also worth noticing that the clustering of the SDSS (Sloan Digital 
Sky Survey) LRG (Luminous Red Galaxies)  
sample~\cite{Eisensteinetal2005}, for instance, shows a slightly
sharper acoustic peak than expected  from linear theory and smearing
due to nonlinearities. However, one should remember that the data
points are strongly correlated, so  that a very high acoustic peak is
actually allowed by the current $\Lambda$CDM cosmology.  Future
redshift surveys such as ADEPT, BOSS, CIP, DES, HETDEX, LSST,
Pan-STARRS,  PAU, WiggleZ or WFMOS ~\cite{galaxysurveys}, which will
obtain redshifts for millions of  galaxies, should achieve an
exquisite precision on the shape of the  baryon acoustic signature in
the clustering of galaxies.  Beyond the nature of dark energy, a
precise measurement of the BAO  could also place constraints on galaxy
biasing and the physical  mechanisms that cause it.

\acknowledgments

I am indebted to Ravi Sheth for pointing out the inconsistency of the
expression for the peak mean streaming in an earlier version of this
manuscript. I would also like to thank Robert Smith for helpful
comments on a preliminary draft; Ilian Iliev and Uro\u{s} Seljak  for
interesting discussions; Martin Crocce and Teppei Okumura for
stimulating correspondence. This work is supported by the Swiss
National Foundation under contract No. 200021-116696/1.

\appendix

\section{Mean streaming of peak pairs}
\label{sec:pkstreamdetails}

\subsection{Correlations of velocity field}

Let us introduce the scaled velocity field
$\vu(\vq)=\vv(\vq)/(a\beta(a)\sigma_{-1})$. The rms variance
$a\beta\sigma_{-1}$ is the three-dimensional proper velocity
dispersion of random  field points, which is $\sim 430\kms$ at present
time in the cosmology considered  here. Also, the notational shorthand
$\Delta\vu$ will designate  the difference $\vu(\vq_2)-\vu(\vq_1)$.
The auto-correlation of the velocity and its cross-correlations with
the fields $\eta_i$, $\nu$ and $\zeta_{ij}$ can be written as
\begin{eqnarray}
\lefteqn{\la\upsilon_i(\vq_1)\upsilon_j(\vq_2)\ra =  {\cal
U}_1(r)\,\rh_i\rh_j + {\cal U}_2(r)\,\delta_{ij}} && \\
\lefteqn{\la\upsilon_i(\vq_1)\eta_j(\vq_2)\ra =
\gamu\,\xi_1(r)\,\rh_i\rh_j + \gamu\,\xi_2(r)\,\delta_{ij}} &&
\nonumber \\ \lefteqn{\la\nu(\vq_1)\upsilon_i(\vq_2)\ra = {\cal
V}(r)\,\rh_i} && \nonumber \\
\lefteqn{\la\upsilon_i(\vq_1)\zeta_{lm}(\vq_2)\ra} && \nonumber \\
&& ={\cal S}_1(r)\,\rh_i\rh_l\rh_m +{\cal
S}_2(r)\,\left(\rh_i\delta_{lm}  + \rh_l\delta_{im} +
\rh_m\delta_{il}\right) \nonumber\;.
\end{eqnarray} 
Here $\upsilon_i(\vq)$ designates the components of $\vu(\vq)$.
Notice that ${\cal U}_\parallel(r)={\cal U}_1(r)+{\cal U}_2(r)$ and
${\cal U}_\perp(r)={\cal U}_2(r)$ are the radial and transverse
correlation functions of the velocity field~\cite{corrvel}.  For sake
of completeness, the various angle average correlations are
\begin{eqnarray}
{\cal U}_1(r)\!\!\! &=& \!\!\!
-\frac{1}{\sigma_{-1}^2}\int_0^\infty\!\!  \dd\ln k\,
k^{-2}\Delta^2(k)\,j_2(kr) \\ {\cal U}_2(r)\!\!\! &=& \!\!\!
\frac{1}{\sigma_{-1}^2}\int_0^\infty\!\!  \dd\ln k\,
k^{-2}\Delta^2(k)\,\left[\frac{1}{3}j_0(kr)+\frac{1}{3}j_2(kr) \right]
\nonumber \\ \xi_1(r)\!\!\! &=& \!\!\!
-\frac{1}{\sigma_0^2}\int_0^\infty\!\!  \dd\ln k\,
\Delta^2(k)\,j_2(kr) \nonumber \\ \xi_2(r)\!\!\! &=& \!\!\!
\frac{1}{\sigma_0^2}\int_0^\infty\!\!  \dd\ln k\,
\Delta^2(k)\,\left[\frac{1}{3}j_0(kr)+\frac{1}{3}j_2(kr) \right]
\nonumber \\ {\cal V}(r)\!\!\! &=& \!\!\!
-\frac{1}{\sigma_{-1}\sigma_0}\int_0^\infty\!\!  \dd\ln k\,
k^{-1}\Delta^2(k)\,j_1(kr) \nonumber \\ {\cal S}_1(r)\!\!\! &=& \!\!\!
-\frac{1}{\sigma_{-1}\sigma_2}\int_0^\infty\!\!  \dd\ln k\,
k\Delta^2(k)\,j_3(kr) \nonumber \\ {\cal S}_2(r)\!\!\! &=& \!\!\!
\frac{1}{\sigma_{-1}\sigma_2}\int_0^\infty\!\!  \dd\ln k\,
k\Delta^2(k)\,\left[\frac{1}{5}j_1(kr)+\frac{1}{5}j_3(kr) \right]
\nonumber
\label{eq:correlv}
\end{eqnarray}
for the Gaussian density field considered here. The functions $\xi_1$
and $\xi_2$ satisfy the relation $\xi_1(r)+3\,\xi_2(r)=\xi(r)$. Like
the spectral width $\gamma$, the parameter
$\gamu=\sigma_0^2/(\sigma_{-1}\sigma_1)$ characterises the range over
which the velocity power spectrum $\propto k^{-2}\Delta^2(k)$ is
large. It should be noted that the latter peaks on scale much
larger than the density  power spectrum. Also, the correlation ${\cal
V}(r)$ is proportional to  the mean streaming of ambient field points,
\begin{equation}
\la\left[1+\delta(\vq_1)\right]\left[1+\delta(\vq_2)\right]
\Delta\vu\cdot\rvh\ra=2\sigma_0{\cal V}(r)\;,
\end{equation}
which is mass weighted by the densities at $\vq_1$ and $\vq_2$.

\subsection{Mean streaming at leading order}

The calculation of the peak pairwise velocity is more intricate than
the peak correlation since we have three additional degrees of freedom, 
but it closely follows the analysis described in Sec.~\ref{sec:pkcorr}. 

The line of sight pairwise velocity weighted over all pairs with 
comoving separation $r$ can be expressed as
\begin{eqnarray}
\lefteqn{\left[1+\xpk(r)\right]u_{12}(r)=\la\npk\ra^{-2}} && \\ 
&& \times \frac{1}{4\pi}\int\!\!\dd\Omega_{\rvh}\,\dd\vy_1\dd\vy_2
\left(\Delta\vu\cdot\rvh\right)\npk(\vq_1)\npk(\vq_2)\,
P(\vy_1,\vy_2;\vr) \nonumber 
\label{eq:pairwise}
\end{eqnarray}
The local peak density $\npk(\vq)$ is given by
equation~(\ref{eq:next}), supplemented by the appropriate  conditions
to select those maxima with a certain threshold height.  To obtain the
average pair velocity as a function of separation $r$, we  need to
calculate the 2-point probability distribution for the variables
$\vy^\top=(\upsilon_i,\eta_i,\nu,\zeta_A)$. At zero lag, both $v_i$ 
and $\eta_i$ are uncorrelated with the density and the Hessian
$\zeta_A$. Hence,  the covariance $\vmm_1$ of the components
$(\upsilon_i,\eta_i,\nu)$ is a  $7\times 7$ block matrix which reads
\begin{equation}
\vmm_1=\left(\begin{array}{ccc} 1/3\,\vii & \gamu/3\,\vii & 
0_{3\times 1} \\
\gamu/3\,\vii & 1/3\,\vii & 0_{3\times 1} \\ 
0_{1\times 3} & 0_{1\times 3} & 1 \end{array}\right)\;.
\end{equation}
Similarly, the covariance $\vmm_2$ of the Hessian, and the 
cross-covariance $\vmm_3$ between $\zeta_A$ and the entries 
$(v_i,\eta_i,\nu)$ are
\begin{eqnarray}
\vmm_2 &=& \left(\begin{array}{cc}\vaa/15 & 0_{3\times 3} \\ 
0_{3\times 3} & \vii/15 \end{array}\right) \nonumber \\ 
\vmm_3 &=& \left(\begin{array}{ccc} 0_{3\times 3} & 0_{3\times 3} & 
-\gamma/3\,1_{3\times 1} \\ 0_{3\times 3} & 0_{3\times 3} & 0_{3\times 1}
\end{array}\right)\;.
\end{eqnarray}
Proceeding as in Sec.~\ref{sec:pkcorr}, we now consider the regime where 
all the correlations are much less than unity. The 2-point probability 
distribution $P(\vy_1,\vy_2;\vr)$ can thus be expanded in the small 
perturbation $\vbb(\vr)$,
\begin{equation}
P(\vy_1,\vy_2;\vr)\approx \left(1+\vy_1^\top\vmm^{-1}
\vbb\,\vmm^{-1}\vy_2\right)\,P(\vy_1) P(\vy_2)\;.
\end{equation}
Here $P(\vy)$ designates the 1-point probability density.
As before, the (now $13\times 13$) matrix $\vbb(\vr)$ denotes the 
covariances at 
different comoving positions. It has a (unique) harmonic decomposition 
in term of the matrices $\vbb_i^{\ell,m}$ (equation~(\ref{eq:multipole}). 
The computation of these matrices is, however, unnecessary as we will see
later. Furthermore, the quadratic form $\bar{Q}(\vy_1,\vy_2)$ now reads
\begin{eqnarray}
2\bar{Q}&=&\frac{3\upsilon_1^2}{1-\gamu^2}+\nu_1^2
+\frac{\left(\gamma\nu_1+\tr\zeta_1\right)^2}{1-\gamma^2} \nonumber \\
&& +\frac{5}{2}\left[3\tr(\zeta_1^2)-\left(\tr\zeta_1\right)^2\right] 
+ 1\leftrightarrow 2\;.
\label{eq:qform}
\end{eqnarray} 
We note that the velocity dispersion of density maxima is lower by a
factor $1-\gamu^2$ than that of random field
points~\cite{Bardeenetal1986}. One has $\gamu\approx 0.43$ for a
smoothing length $R_f=5\hmpc$. Moreover, eq.~(\ref{eq:qform}) leads to
a one-point  probability distribution $P(\vy|{\rm peak})\propto
\exp[-\bar{Q}(\vy)]$ separable into the product
$P_\upsilon(\upsilon_i)P_{\nu\zeta}(\nu,\zeta_A)$, where
$P_{\nu\zeta}$ is the one-point distribution of the density and its
second derivatives, and $P_\upsilon(\upsilon_i)$ is the velocity
distribution of peaks,
\begin{equation}
P_\upsilon(\upsilon_i|{\rm peak})=\frac{3^{3/2}}
{(2\pi)^{3/2}\left(1-\gamu^2\right)^{3/2}}\,
\exp\left[-\frac{3\upsilon^2}{2\left(1-\gamu^2\right)}\right]\;.
\end{equation}
The separability of the one-point distribution separability considerably 
simplifies the calculation.

Taking the product $(\Delta\vu\cdot\rvh)\,\vbb(\vr)$ mixes the various
multipole matrices $\vbb_i^{\ell,m}$, so that the result depends on
the correlation functions of $\upsilon_i$, $\eta_i$, $\nu$ and
$\zeta_A$  in a rather complicated way. Averaging over the directions
gives
\begin{equation}
\tilde{\vbb}=
\frac{1}{4\pi}\int\!\!\dd\Omega_{\rvh}\,\left(\Delta\vu\cdot\rvh\right)
\,\vmm^{-1}\vbb\,\vmm^{-1} = 
\left(\begin{array}{cc}\tilde{\vbb}_1 & -\tilde{\vbb}_3^\top \\ 
\tilde{\vbb}_3 & \tilde{\vbb}_2 \end{array}\right)\;,
\label{eq:tvbb}
\end{equation}
where the block matrices $\tilde{\vbb}_i$ have the same dimensions as 
$\vmm_i$. The minus sign in the right-hand side of eq.~\ref{eq:tvbb}
arises from the negative parity of the correlations 
$\la\upsilon_i\,\zeta_{lm}\ra$ and $\la\eta_i\,\zeta_{lm}\ra$.
Owing to the angular average, 
the calculation of the $\vbb_i^{\ell,m}$ can be avoided by writing down 
the entries of $\vbb_i(\vr)$ using the relations eq.~(\ref{eq:correl}) 
and (\ref{eq:correlv}), and retaining only those components involving 
odd products of the unit vector $\rvh_i$.
A tedious calculation shows that $\tilde{\vbb}_1(r)$ and  
$\tilde{\vbb}_3(r)$ can be cast into the form
\begin{eqnarray}
\tilde{\vbb}_1 &=& \left(\begin{array}{ccc} 
0_{3\times 3} & 0_{3\times 3} & -\alpha_1\Delta\vu \\ 
0_{3\times 3} & 0_{3\times 3} & -\alpha_2\Delta\vu \\ 
\alpha_1\Delta\vu^\top & \alpha_2\Delta\vu^\top & 0 
\end{array}\right) \\
\tilde{\vbb}_3 &=& \left(\begin{array}{ccc}
\gamma\alpha_1\Upsilon_1-3\alpha_3\Upsilon_2 & 
\gamma\alpha_2\Upsilon_1-3\alpha_4\Upsilon_2 & 0_{3\times 1} \\
-3\alpha_3\Upsilon_3 & -3\alpha_4\Upsilon_3 & 0_{3\times 1} 
\end{array}\right) \nonumber \;,
\end{eqnarray} 
The functions $\alpha_i(r)$ are
\begin{eqnarray}
\alpha_1(r) &=& 
\frac{{\cal V}-\gamu\Xi-
\gamma\left({\cal S}-\gamu\Pi\right)}{\left(1-\gamma^2\right)
\left(1-\gamu^2\right)} \nonumber \\
\alpha_2(r) &=&
\frac{\Xi-\gamu{\cal V}-
\gamma\left(\Pi-\gamu{\cal S}\right)}{\left(1-\gamma^2\right)
\left(1-\gamu^2\right)} \nonumber \\
\alpha_3(r) &=& \frac{{\cal S}-\gamu\Pi}{1-\gamu^2} \nonumber \\
\alpha_4(r) &=& \frac{\Pi-\gamu{\cal S}}{1-\gamu^2} \;,
\end{eqnarray}
where we have omitted the explicit $r$-dependence of the correlations 
for brevity. The $3\times 3$ matrices $\Upsilon_i$ have the components 
$\Delta\upsilon_i$ of the vector $\Delta\vu$ as entries,
\begin{eqnarray}
\Upsilon_1 &=& \left(\begin{array}{ccc} 
\Delta\upsilon_1 & \Delta\upsilon_2 & \Delta\upsilon_3 \\ 
\Delta\upsilon_1 & \Delta\upsilon_2 & \Delta\upsilon_3 \\ 
\Delta\upsilon_1 & \Delta\upsilon_2 & \Delta\upsilon_3 
\end{array}\right) \nonumber \\
\Upsilon_2 &=& \left(\begin{array}{ccc} 
\Delta\upsilon_1 & 0 & 0 \\
0 & \Delta\upsilon_2 & 0 \\
0 & 0 & \Delta\upsilon_3 
\end{array}\right) \nonumber \\
\Upsilon_3 &=& \left(\begin{array}{ccc} 
\Delta\upsilon_2 & \Delta\upsilon_1 & 0 \\
\Delta\upsilon_3 & 0 & \Delta\upsilon_1 \\
0 & \Delta\upsilon_3 & \Delta\upsilon_2 
\end{array}\right)\;.
\end{eqnarray}
We have also set
\begin{equation}
\Pi(r)=\Pi_1+5\Pi_2,~~~
{\cal S}(r)={\cal S}_1+5{\cal S}_2\;.
\end{equation}
The matrix $\tilde{\vbb}_2$ is identically zero. 

The rest of the  calculation is easily accomplished owing to the
factorisation of the  one-point probability distribution $P(\vy|{\rm
peak})$.  Notice that the  scalar $\vy_1^\top\tilde{\vbb}\vy_2$ contains
terms linear and quadratic  in $\vu_1$ and $\vu_2$. After integrating
out the velocities, the  linear terms vanish and we eventually find
\begin{eqnarray}
\lefteqn{\int\!\!\dd^3\vu_1\dd^3\vu_2\,\vy_1^\top\tilde{\vbb}\vy_2\,
P(\vu_1|{\rm peak})P(\vu_2|{\rm peak})=} && \\
&& \left[\alpha_1\left(\nu_1+\nu_2\right)+
\left(\gamma\alpha_1-\alpha_3\right)\left(\tr\zeta_1+\tr\zeta_2\right)
\right]\left(1-\gamu^2\right) \nonumber \;. 
\end{eqnarray}
Transforming to the set of variables $(u_i,v_i,w_i)$ and substituting 
the expressions~(\ref{eq:biases}) of the bias parameters $b_\nu$ and 
$b_\zeta$, the mean streaming of peak pairs can be recast into the 
form of equation~(\ref{eq:pkstream}).


\begin{thebibliography}{50}

\bibitem[\protect\citeauthoryear{}{}]{baotheory}{
P.J.E.~Peebles, J.T.~Yu, \ApJ, {\bf 162}, 815 (1970);
R.~Sunyaev, Ya.-B~Zeldovich, \ApJSS, {\bf 7}, 3 (1970);
J.R.~Bond, G.~Efstathiou, \ApJ, {\bf 285}, L45 (1984);
J.A.~Holtzmann, \ApJS, {\bf 71}, 1 (1989);
W.~Hu, N.~Sugiyama, \ApJ, {\bf 471}, 542 (1996);
D.J.~Eisenstein, W.~Hu, \ApJ, {\bf 496}, 605 (1998).
\label{baotheory}}

\bibitem[\protect\citeauthoryear{Eisenstein et al.}{2005}]
{Eisensteinetal2005} D.J.~Eisenstein, et al., \ApJ, {\bf 633}, 560 
(2005).

\bibitem[\protect\citeauthoryear{}{}]{baoobs}{
S.~Cole, et al., \MNRAS, {\bf 362}, 505 (2005);
G.~H\"utsi, \AA, {\bf 449}, 891 (2006);
M.~Tegmark, et al., \PRD, {\bf }74, 123507 (2006);
W.J.~Percival, et al., \ApJ, {\bf 657}, 51 (2007);
W.J.~Percival, et al., \MNRAS, {\bf 381}, 1053 (2007);
C.~Blake, A.~Collister, S.~Bridle, O.~Lahav, \MNRAS, {\bf 374}, 
1527 (2007); 
N.~Padmanabhan, et al., \MNRAS, {\bf 378}, 852 (2007);
T.~Okumura, et al., \ApJ, {\bf 676}, 889 (2008).
\label{baoobs}}

\bibitem[\protect\citeauthoryear{Estrada et al.}{2008}]{Estradaetal2008}
J.~Estrada, E.~Sefusatti, J.A.~Frieman, ArXiv Astrophysics 
eprint:astro-ph/0801.3485 (2008).

\bibitem[\protect\citeauthoryear{}{}]{baoprobe}{
W.~Hu, M.~White, \ApJ, {\bf 471}, 30 (1996);
D.J.~Eisenstein, W.~Hu, M.~Tegmark, \ApJ, {\bf 504}, L57 (1998);
A.~Cooray, W.~Hu, D.~Huterer, M.~Joffre, \ApJ, {\bf 557}, L7 (2001);
W.~Hu, Z.~Haiman, \PRD, {\bf 68}, 063004 (2003);
C.~Blake, K.~Glazebrook, \ApJ, {\bf 594}, 665 (2003);
E.V.~Linder, \PRD, {\bf 68}, 3504 (2003);
T.~Matsubara, \ApJ, {\bf 615}, 573 (2004);
L.~Amendola, C.~Quercellini, E.~Giallongo, \MNRAS, {\bf 357}, 429 (2005);
C.~Blake, S.~Bridle, \MNRAS, {\bf 363}, 1329 (2005);
K.~Glazebrook, C.~Blake, \ApJ, {\bf 631}, 1 (2005);
D.~Dolney, B.~Jain, M.~Takada, \MNRAS, {\bf 366}, 884 (2006);
H.~Zhan, L.~Knox, \ApJ, {\bf 644}, 663 (2006);
C.~Blake, et al., \MNRAS, {\bf 365}, 255 (2006);
N.~Padmanabhan, M.~White, ArXiv Astrophysics eprint:astro-ph/0804.0799 
(2008);
M.~Shoji, D.~Jeong, E.~Komatsu, ArXiv Astrophysics eprint:astro-ph/0805.4238
(2008).
\label{baoprobe}}

\bibitem[\protect\citeauthoryear{}{}]{3dbias}{
Magnification bias and, to a lesser extent, stochastic deflections also
affect baryonic features in the 3D galaxy correlation. See, for instance,
T.~Nishimichi, et al., \PASJ, {\bf 59}, 93 (2007); L.~Hui, E.~Gazta\~naga,
M.~Loverde, \PRD, {\bf 76}, 103502 (2007); 
A.~Vallinotto, S.~Dodelson, C.~Schimd, J.-P.~Uzan, \PRD, {\bf 75}, 103509 
(2007); 
M.~Loverde, L.~Hui, E.~Gazta\~naga, \PRD, {\bf 77}, 023512 (2008);
L.~Hui, E.~Gazta\~naga, M.~Loverde, \PRD, {\bf 77}, 063526 (2008).
\label{3dbias}}

\bibitem[\protect\citeauthoryear{}{}]{baosimu}{
H-J.~Seo, D.J.~Eisenstein, \ApJ, {\bf 598}, 720 (2003);
M.~White, \AstroPart, {\bf 24}, 334 (2005);
V.~Springel, et al., \Nat, {\bf 435}, 629 (2005);
R.E.~Angulo, et al., \MNRAS, {\bf 362}, L25 (2005).
E.~Huff, A.E.~Schultz, M.~White, D.J.~Schlegel, M.S.~Warren, \AstroPart,
{\bf 26}, 351 (2007);
Z.~Ma, \ApJ, {\bf 665}, 887 (2007);
R.E.~Angulo, C.M.~Baugh, C.S.~Frenk, C.G.~Lacey, \MNRAS, {\bf 383}, 755
(2008);
R.~Takahashi, et al., ArXiv Astrophysics eprint:astro-ph/0802.1808 (2008);
A.G.~Sanchez, C.M.~Baugh, R.~Angulo, ArXiv Astrophysics 
eprint:astro-ph/0804.0233 (2008).
\label{baosimu}}

\bibitem[\protect\citeauthoryear{Meiksin et al.}{1999}]{Meiksinetal1999}
A.~Meiksin, M.~White, J.A.~Peacock, \MNRAS, {\bf 304}, 851 (1999).

\bibitem[\protect\citeauthoryear{Seo \& Eisenstein}{2005}]{SeoEisenstein2005}
H.J.-Seo, D.J.~Eisenstein, \ApJ, {\bf 633}, 575 (2005).

\bibitem[\protect\citeauthoryear{Jeong \& Komatsu}{2006}]{JeongKomatsu2006}
D.~Jeong, E.~Komatsu, \ApJ, {\bf 651}, 619 (2006).

\bibitem[\protect\citeauthoryear{Schultz \& White}{2006}]{SchultzWhite2006}
A.E.~Schultz, M.~White, \AstroPart, {\bf 25}, 172 (2006).

\bibitem[\protect\citeauthoryear{Guzik et al.}{2007}]{Guziketal2007}
J.~Guzik, G.~Bernstein, R.E.~Smith, \MNRAS, {\bf 375}, 1329 (2007).

\bibitem[\protect\citeauthoryear{Eisenstein et al.}{2007}]
{Eisensteinetal2007a} D.J.~Eisenstein, H.-J.~Seo, M.~White, \ApJ, 
{\bf 664}, 660 (2007).

\bibitem[\protect\citeauthoryear{Smith et al.}{2007}]{Smithetal2007}
R.E.~Smith, R.~Soccimarro, R.K.~Sheth, \PRD, {\bf 75}, 063512 (2007).

\bibitem[\protect\citeauthoryear{Matarrese \& Pietroni}{2007}]
{MatarresePietroni2007} S.~Matarrese, M.~Pietroni, \JCAP, {\bf 06}, 026 
(2007).

\bibitem[\protect\citeauthoryear{Crocce \& Scoccimarro}{2008}]
{CrocceScoccimarro2008} M.~Crocce, R.Scoccimarro, \PRD, {\bf 77}, 023533 
(2008).

\bibitem[\protect\citeauthoryear{Smith et al.}{2008}]{Smithetal2008}
R.E.~Smith, R.Scoccimarro, R.K.~Sheth, \PRD, {\bf 77}, 043525 (2008).

\bibitem[\protect\citeauthoryear{Matsubara}{2008}]{Matsubara2008}
T.~Matsubara, \PRD, {\bf 77}, 063530 (2008).

\bibitem[\protect\citeauthoryear{Matarrese \& Pietroni}{2008}]
{MatarresePietroni2008} S.~Matarrese, M.~Pietroni, \MPLA, {\bf 23},
25 (2008).

\bibitem[\protect\citeauthoryear{Seo et al.}{2008}]{Seoetal2008}
H.-J.~Seo, E.R.~Siegel, D.J.~Eisenstein, M.~White, ArXiv Astrophysics 
eprint:astro-ph/0805.0117 (2008). 

\bibitem[\protect\citeauthoryear{}{}]{JeongKomatsu2008}
D.~Jeong, E.~Komatsu, ArXiv Astrophysics eprint:astro-ph/0805.2632

\bibitem[\protect\citeauthoryear{Eisenstein et al.}{2007}]
{Eisensteinetal2007b} D.J.~Eisenstein, H.-J.~Seo, E.~Sirko, D.N.~Spergel, 
\ApJ, {\bf 664}, 675 (2007).

\bibitem[\protect\citeauthoryear{Coles}{1993}]{Coles1993} P.~Coles,
\MNRAS, {\bf 262}, 1065 (1993).

\bibitem[\protect\citeauthoryear{Scherrer \& Weinberg}{1998}]
{ScherrerWeinberg1998} R.J.~Scherrer, D.H.~Weinberg, \ApJ, {\bf 504},
607 (1998).

\bibitem[\protect\citeauthoryear{Fry \& Gazta\~{n}aga}{1993}]
{FryGaztanaga1993} J.N.~Fry, E.~Gazta\~{n}aga, \ApJ, {\bf 413}, 447
(1993).

\bibitem[\protect\citeauthoryear{Szalay}{1988}]{Szalay1988}
A.S.~Szalay, \ApJ, 1988, {\bf 333}, 21.

\bibitem[\protect\citeauthoryear{}{}]{gaussianity}{
Komatsu, et al., \ApJS, {\bf 148}, 119 (2003);
P.~Creminelli, N.~Alberto, L.~Senatore, M.~Tegmark, M.~Zaldarriaga,
\JCAP, {\bf 06}, 004 (2006);
A.~Yadav, B.D.~Wandelt, ArXiv Astrophysics eprint:astro-ph/0712.1148 (2007);
A.~Slosar, et al., ArXiv Astrophysics eprint:astro-ph/0805.3580 (2008).
\label{gaussianity}}

\bibitem[\protect\citeauthoryear{}{}]{wmap5}{ 
E.~Komatsu, et al., ArXiv Astrophysics eprint:astro-ph/0803.0547 (2008);
E.L.~Wright, et al., ArXiv Astrophysics eprint:astro-ph/0803.0577 (2008);
J.~Dunkley, et al., ArXiv Astrophysics eprint:astro-ph/0803.0586 (2008);
M.R.~Nolta, et al., ArXiv Astrophysics eprint:astro-ph/0803.0593 (2008);
B.~Gold, et al., ArXiv Astrophysics eprint:astro-ph/0803.0715 (2008);
G.~Hinshaw, et al., ArXiv Astrophysics eprint:astro-ph/0803.0732 (2008). 
\label{wmap5}}

\bibitem[\protect\citeauthoryear{Bardeen et al.}{1986}]{Bardeenetal1986} 
J.M.~Bardeen, J.R.~Bond, N.~Kaiser, A.S.~Szalay, \ApJ, {\bf 304}, 15 
(1986). 

\bibitem[\protect\citeauthoryear{Kaiser}{1984}]{Kaiser1984} N.~Kaiser,
\ApJ, {\bf 284}, L9 (1984).

\bibitem[\protect\citeauthoryear{Politzer \& Wise}{1984}]
{PolitzerWise1984} D.~Politzer, M.~Wise, \ApJ, {\bf 285}, L1 (1984).

\bibitem[\protect\citeauthoryear{Jensen \& Szalay}{1986}]
{JensenSzalay1986} L.G.~Jensen, A.S.~Szalay, \ApJ, {\bf 305}, L5 (1986).

\bibitem[\protect\citeauthoryear{Doroshkevich}{1970}]{Doroshkevich1970} 
A.G.~Doroshkevich, Astrofizika, {\bf 3}, 175 (1970).

\bibitem[\protect\citeauthoryear{Hoffman \& Shaham}{1985}]{HoffmanShaham1985}
Y.~Hoffman, J.~Shaham, \ApJ, {\bf 297}, 16 (1985).

\bibitem[\protect\citeauthoryear{Peacock \& Heavens}{1985}]{PeacockHeavens1985}
J.A.~Peacock, A.F.~Heavens, \MNRAS, {\bf 217}, 805 (1985).

\bibitem[\protect\citeauthoryear{Coles}{1989}]{Coles1989}
P.~Coles, \MNRAS, {\bf 238}, 319 (1989).

\bibitem[\protect\citeauthoryear{Lumsden, Heavens \& Peacock}{1989}]
{Lumsdenetal1989} S.L.~Lumsden, A.F.~Heavens, J.A.~Peacock, \MNRAS,
{\bf 238}, 293 (1989).

\bibitem[\protect\citeauthoryear{Kaiser \& Davis}{1985}]{KaiserDavis1985}
N.~Kaiser, M.~Davis, \ApJ, {\bf 297}, 365 (1985).

\bibitem[\protect\citeauthoryear{Otto et al.}{1986}]{Ottoetal1986}
S.~Otto, H.D.~Politzer, M.B.~Wise, \PRL, {\bf 56}, 2772 (1986).

\bibitem[\protect\citeauthoryear{Cen}{1998}]{Cen1998}
R.~Cen, \ApJ, {\bf 509}, 494 (1998).

\bibitem[\protect\citeauthoryear{Desjacques}{2007}]{Desjacques2007}
V.~Desjacques, \MNRAS, {\bf 388}, 638 (2008).

\bibitem[\protect\citeauthoryear{Desjacques \& Smith}{2008}]
{DesjacquesSmith2008} V.~Desjacques, R.E.~Smith, \PRD, {\bf 78}, 
023527 (2008).

\bibitem[\protect\citeauthoryear{Adler \& Taylor}{2007}]{AdlerTaylor2007}
R.J.~Adler, J.~Taylor, Random Fields and Geometry, Springer Monographs
in Mathematics, Springer, New York (2007).

\bibitem[\protect\citeauthoryear{}{}]{boltzmann}{
U.~Seljak, M.~Zaldarriaga, \ApJ, {\bf 469}, 437 (1996);
A.~Lewis, A.~Challinor, A.~Lasenby, \ApJ, {\bf 538}, L473 (2000);
\label{boltzmann}}

\bibitem[\protect\citeauthoryear{}{}]{tksmalltail}{
W.~Hu, N.~Sugiyama, \ApJ, {\bf 471}, 542 (1996);
K.~Yamamoto, N.~Sugiyama, H.~Sato, \PRD, {\bf 56}, 7566 (1997);
K.~Yamamoto, N.~Sugiyama, H.~Sato, \ApJ, {\bf 501}, 442 (1998);
S.~Naoz, R.~Barkana, \MNRAS, {\bf 362}, 1047 (2005).
\label{tksmalltail}}

\bibitem[\protect\citeauthoryear{Kac}{1943}]{Kac1943} M.~Kac, \BAMS, 
{\bf 49}, 314 (1943).

\bibitem[\protect\citeauthoryear{Rice}{1954}]{Rice1954} S.O.~Rice, 
Mathematical analysis of random noise, in Selected Papers on Noise
and Stochastic Processes, Dover, New York (1954). 

\bibitem[\protect\citeauthoryear{Belyaev}{1967}]{Belyaev1967} 
J.K.~Belyaev, Sov.~Math.~Dokl., {\bf 8}, 1107 (1967).

\bibitem[\protect\citeauthoryear{Cline et al.}{1987}]{Clineetal1987}
J.M.~Cline, H.D.~Politzer, S.-Y.~Rey, M.B.~Wise, \CMP, {\bf 112}, 
217 (1987).

\bibitem[\protect\citeauthoryear{Bond et al.}{1991}]{Bondetal1991}
J.R.~Bond, S.~Cole, G.~Efstathiou, N.~Kaiser, \ApJ, {\bf 379}, 440
(1991).

\bibitem[\protect\citeauthoryear{Press \& Schechter}{1974}]
{PressSchechter1974} W.H.~Press, P.~Schechter, \ApJ, {\bf 187}, 425 (1974).

\bibitem[\protect\citeauthoryear{Gunn \& Gott}{1972}]{GunnGott1972} 
J.E.~Gunn, J.R.~Gott III, \ApJ, {\bf 176}, 1 (1972).

\bibitem[\protect\citeauthoryear{Mo \& White}{1996}]{MoWhite1996}
H.J.~Mo, S.D.M.~White, \MNRAS, {\bf 282}, 347 (1996).

\bibitem[\protect\citeauthoryear{Sheth \& Tormen}{1999}]{ShethTormen1999}
R.K~Sheth, G.~Tormen, \MNRAS, {\bf 308}, 119 (1999).

\bibitem[\protect\citeauthoryear{Cole \& Kaiser}{1989}]{ColeKaiser1989}
S.~Cole, N.~Kaiser, \MNRAS, {\bf 237}, 1127 (1989).

\bibitem[\protect\citeauthoryear{Sheth \& Tormen}{2002}]{ShethTormen2002}
R.K.~Sheth, G.~Tormen, \MNRAS, {\bf 329}, 61 (2002).

\bibitem[\protect\citeauthoryear{Kauffmann et al.}{1999}]{Kauffmannetal1999}
G.~Kauffmann, J.M.~Colberg, A.~Diafero, S.D.M.~White, \MNRAS, {\bf 303},
188 (1999).

\bibitem[\protect\citeauthoryear{}{}]{highmassend}{
D.~Reed, et al., \MNRAS, {\bf 346}, 565 (2003);
K.~Heitmann, Z.~Luki\'c, S.~Habib, P.M.~Ricker, \ApJ, {\bf 646}, L1 
(2006);
I.T.~Iliev, et al., \MNRAS, {\bf 369}, 1625 (2006); 
D.~Reed, R.~Bower, C.S.~Frenk, A.~Jenkins, T.~Theuns, \MNRAS, {\bf 374},
2 (2007);
Z.~Luki\'c, K.~Heitmann, S.~Habib, S.~Bashinsky, P.M.~Ricker, \ApJ, 
{\bf 671}, 1160 (2007);
O.~Zahn, et al., \ApJ, {\bf 654}, 12 (2007).
\label{highmassend}}

\bibitem[\protect\citeauthoryear{Bond et al.}{1980}]{Bondetal1980}
J.R.~Bond, G.~Efstathiou, J.~Silk, \PRL, {\bf 45}, 1980 (1980).

\bibitem[\protect\citeauthoryear{}{}]{tkwdm}{
P.~Bode, J.P.~Ostriker, N.~Turok, \ApJ, {\bf 556}, 93 (2001);
S.H.~Hansen, J.~Lesgourgues, S.~Pastor, J.~Silk, \MNRAS, {\bf 333}, 544
(2002).
\label{tkwdm}}

\bibitem[\protect\citeauthoryear{Dalal et al.}{2008}]{Dalaletal2008}
N.~Dalal, M.~White, J.R.~Bond, A.~Shirokov, ArXiv Astrophysics
eprint:astro-ph/0803.3453 (2008).

\bibitem[\protect\citeauthoryear{Zeldovich}{1970}]{Zeldovich1970} 
Y.B.~Zeldovich, \AA, 5, 84 (1970).

\bibitem[\protect\citeauthoryear{Peebles}{1980}]{Peebles1980}
P.J.E~Peebles, The Large-Scale Structure of the Universe (Princeton
University Press, 1980).

\bibitem[\protect\citeauthoryear{Grinstein \& Wise}{1987}]
{GrinsteinWise1987} B.~Grinstein, M.B.~Wise, \ApJ, {\bf 320}, 448 (1987).

\bibitem[\protect\citeauthoryear{}{}]{zapeak}{
R.G.~Mann, A.F.~Heavens, J.A.~Peacock, \MNRAS, {\bf 263}, 798 (1993);
S.~Borgani, P.~Coles, L.~Moscardini, \MNRAS, {\bf 271}, 223 (1994).
\label{zapeak}}

\bibitem[\protect\citeauthoryear{}{}]{peakpatch}{
J.R.~Bond, S.T.~Myers, \ApJS, {\bf 103}, 1 (1996);
J.R.~Bond, S.T.~Myers, \ApJS, {\bf 103}, 41 (1996);
J.R.~Bond, S.T.~Myers, \ApJS, {\bf 103}, 63 (1996).
\label{peakpatch}}

\bibitem[\protect\citeauthoryear{Peebles}{1976}]{Peebles1976}
P.J.E.~Peebles, \ApSS, {\bf 45}, 3 (1976).

\bibitem[\protect\citeauthoryear{Davis \& Peebles}{1977}]{DavisPeebles1977}
M.~Davis, P.J.E.~Peebles, \ApJS, {\bf 34}, 425 (1977). 

\bibitem[\protect\citeauthoryear{Sheth et al.}{2001}]{Shethetal2001}
R.K.~Sheth, A.~Diafero, L.~Hui, R.~Scoccimarro, \MNRAS, {\bf 326}, 463
(2001).

\bibitem[\protect\citeauthoryear{}{}]{halomodel}{
A.J.~Benson, S.~Cole, C.S.~Frenk, C.~Baugh, C.~Lacey, \MNRAS, {\bf 311},
793 (2000);
J.A.~Peacock, R.E.~Smith, \MNRAS, {\bf 318}, 1144 (2000);
U.~Seljak, \MNRAS, {\bf 318}, 203 (2000);
R.~Scoccimarro, R.K.~Sheth, L.~Hui, B.~Jain, \ApJ, {\bf 546}, 20 (2001);
A.~Berlind, D.~Weinberg, \ApJ, {\bf 575}, 587 (2002). 
\label{halomodel}}

\bibitem[\protect\citeauthoryear{}{}]{galaxypeak}{
C.S.~Frenk, S.D.M.~White, M.~Davis, G.~Efstathiou, \ApJ, {\bf 327}, 507
(1988);
N.~Katz, T.~Quinn, J.M.~Gelb, \MNRAS, {\bf 265}, 689 (1993);
C.~Porciani, A.~Dekel, Y.~Hoffman, MNRAS, {\bf 332}, 325 (2002).
\label{galaxypeak}}

\bibitem[\protect\citeauthoryear{}{}]{resmith}{
Robert Smith, private communication.
\label{resmith}}

\bibitem[\protect\citeauthoryear{}{}]{galaxysurveys}{
ADEPT:~http://www.jhu.edu/news\_info/news/home06/
aug06/adept.html; 
BOSS:~http://www.sdss3.org/cos\-mo\-logy.php;
CIP:~http://cfa-www.harvard.edu/cip; 
DES:~http://www.darkenergysurvey.org;
HETDEX:~http://www.as.utexas.edu/hetdex;
LSST:~H.~Zhan, L.~Knox, T.J.~Anthony, V.~Margoniner, \ApJ, {\bf 640}, 8 
(2006) (http://www.lsst.org);
Pan-STARRS:~http://pan-starrs.ifa.hawaii.edu;
PAU: N.~Benitez et al., ArXiv Astrophysics eprint~:astro-ph/0807.0535;
WiggleZ:~http://astronomy.swin.edu.au;
WFMOS~: Glazebrook et al., ArXiv Astrophysics eprint~: astro-ph/0507457
(2005).
\label{galaxysurveys}}

\bibitem[\protect\citeauthoryear{}{}]{corrvel}{
A.S.~Monin, A.M.~Yaglom, Statistical Fluid Mechanics (Cambridge MIT
Press, 1975);
K.~G\'orski, \ApJ, {\bf 332}, L7 (1988). 
\label{corrvel}}

\end{thebibliography}
\end{document}